\documentclass[8.5pt,twoside,twocolumn]{article}
\usepackage[super,sort&compress,comma]{natbib} 
\usepackage{mhchem}
\usepackage{times,mathptmx}
\usepackage{sectsty}
\usepackage{balance} 
\usepackage{color}

\usepackage{graphicx} 
\usepackage{lastpage}
\usepackage[format=plain,justification=raggedright,singlelinecheck=false,font=small,labelfont=bf,labelsep=space]{caption} 
\usepackage{fancyhdr}
\begin{document}

\thispagestyle{plain}
\fancypagestyle{plain}{
\fancyhead[L]{\includegraphics[height=8pt]{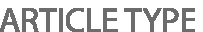}}
\fancyhead[C]{\hspace{-1cm}\includegraphics[height=20pt]{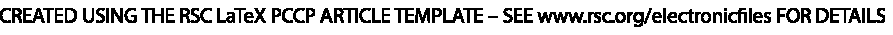}}
\fancyhead[R]{\includegraphics[height=10pt]{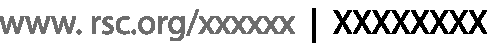}\vspace{-0.2cm}}
\renewcommand{\headrulewidth}{1pt}}
\renewcommand{\thefootnote}{\fnsymbol{footnote}}
\renewcommand\footnoterule{\vspace*{1pt}%
\hrule width 3.4in height 0.4pt \vspace*{5pt}} 
\setcounter{secnumdepth}{5}

\makeatletter 
\def\subsubsection{\@startsection{subsubsection}{3}{10pt}{-1.25ex plus -1ex minus -.1ex}{0ex plus 0ex}{\normalsize\bf}} 
\def\paragraph{\@startsection{paragraph}{4}{10pt}{-1.25ex plus -1ex minus -.1ex}{0ex plus 0ex}{\normalsize\textit}} 
\renewcommand\@biblabel[1]{#1}            
\renewcommand\@makefntext[1]%
{\noindent\makebox[0pt][r]{\@thefnmark\,}#1}
\makeatother 
\renewcommand{\figurename}{\small{Fig.}~}
\sectionfont{\large}
\subsectionfont{\normalsize} 

\fancyfoot{}
\fancyfoot[LO,RE]{\vspace{-7pt}\includegraphics[height=9pt]{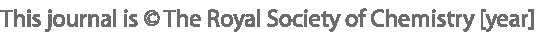}}
\fancyfoot[CO]{\vspace{-7.2pt}\hspace{12.2cm}\includegraphics{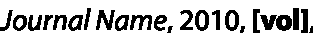}}
\fancyfoot[CE]{\vspace{-7.5pt}\hspace{-13.5cm}\includegraphics{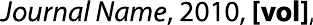}}
\fancyfoot[RO]{\footnotesize{\sffamily{1--\pageref{LastPage} ~\textbar  \hspace{2pt}\thepage}}}
\fancyfoot[LE]{\footnotesize{\sffamily{\thepage~\textbar\hspace{3.45cm} 1--\pageref{LastPage}}}}
\fancyhead{}
\renewcommand{\headrulewidth}{1pt} 
\renewcommand{\footrulewidth}{1pt}
\setlength{\arrayrulewidth}{1pt}
\setlength{\columnsep}{6.5mm}
\setlength\bibsep{1pt}

\twocolumn[
  \begin{@twocolumnfalse}
\noindent\LARGE{\textbf{Negative linear compressibility$^\dag$}}
\vspace{0.6cm}

\noindent\large{\textbf{Andrew B. Cairns and
Andrew L. Goodwin$^{\ast}$}}\vspace{0.5cm}

\noindent\textit{\small{\textbf{Received Xth XXXXXXXXXX 20XX, Accepted Xth XXXXXXXXX 20XX\newline
First published on the web Xth XXXXXXXXXX 200X}}}

\noindent \textbf{\small{DOI: 10.1039/b000000x}}
\vspace{0.6cm}

\noindent \normalsize{While all materials reduce their intrinsic volume under hydrostatic (uniform) compression, a select few actually \emph{expand} along one or more directions during this process of densification. As rare as it is counterintuitive, such ``negative compressibility'' behaviour has application in the design of pressure sensors, artificial muscles and actuators. The recent discovery of surprisingly strong and persistent negative compressibility effects in a variety of new families of materials has ignited the field. Here we review the phenomenology of negative compressibility in this context of materials diversity, placing particular emphasis on the common structural motifs that recur amongst known examples. Our goal is to present a mechanistic understanding of negative compressibility that will help inform a clear strategy for future materials design. }
\vspace{0.5cm}
 \end{@twocolumnfalse}
  ]

\section{Introduction}

\footnotetext{\dag~Electronic Supplementary Information (ESI) available: [details of any supplementary information available should be included here]. See DOI: 10.1039/b000000x/}


\footnotetext{~Inorganic Chemistry Laboratory, Department of Chemistry, University of Oxford, South Parks Road. OX1 3QR; E-mail: andrew.goodwin@chem.ox.ac.uk}

\begin{figure}
\includegraphics{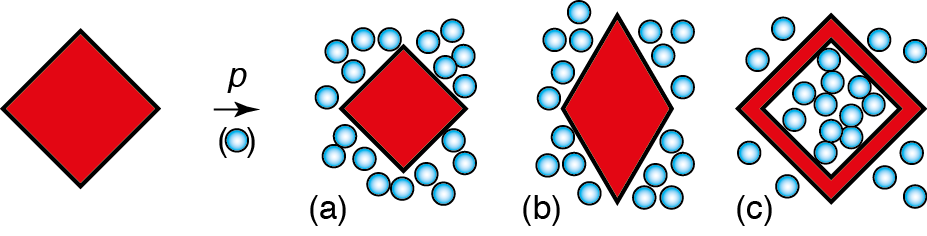}
\caption{Mechanical responses to hydrostatic pressure: (a) positive compressibility---contraction in all directions; (b) negative linear compressibility---linear expansion in one direction; (c) inflation associated with incorporation of the pressure-transmitting media (blue circles) within the material interior. The system volume (represented here by the solid red area) is reduced in all cases.}
\label{fig1}
\end{figure}

Negative linear compressibility (NLC) is the bizarre materials property whereby a system expands along one direction when compressed uniformly.\cite{Gatt:2008,Baughman:1998} Bizarre, because our intuition is that materials should shrink when squeezed---an intuition that is (rightly) grounded in the thermodynamic requirement that volume be reduced at increased pressure.\cite{Nye:1984,Newnham:2005} Yet NLC does not violate thermodynamics: it simply arises whenever volume reduction can be coupled to linear expansion [Fig.~\ref{fig1}]. In the benchmark review of NLC---now 17 years old---Baughman explains how the phenomenon might eventually be applied in a variety of ways, including the development of artificial muscles and amplification of piezoelectric response for next-generation sensors and actuators.\cite{Baughman:1998} Until recently, there has been relatively little hope of identifying suitable candidates for these applications. The most significant challenges have been the apparent rarity of NLC (Ref.~\citenum{Baughman:1998} reports it to occur in only 13 known materials) and the extreme weakness of the NLC effects exhibited by these materials.

Over the past few years, the field has changed in two important respects. The first is that materials have now been discovered that exhibit orders-of-magnitude stronger NLC effects than the ``classical'' NLC materials reviewed by Baughman.\cite{Baughman:1998} The second advance---which likely reflects the improved accessibility of variable-pressure crystallographic measurements---is that NLC has now been found to occur in a much greater diversity of materials, ranging from dense inorganic oxides\cite{Skelton:1976} and fluorides\cite{Mirgorodsky:1994} to metal--organic frameworks\cite{Li:2012,Cai:2014} and even molecular solids.\cite{Woodall:2013} Consequently we felt it timely to review the phenomenon of NLC once again, placing particular emphasis on the common underlying geometric motifs responsible for NLC in the various materials---both old and new---and in doing so to help inform future materials design.

Our review is organised as follows. We begin with an overview of the theoretical and experimental approaches to understanding, measuring and comparing NLC responses. As part of this discussion we introduce the new measure of ``compressibility capacity'', $\chi_K$. This will play a role in allowing us to compare the NLC responses of very different materials. The bulk of the review concerns the NLC behaviour of known materials, grouped according to the microscopic mechanism responsible for NLC. The summary with which our review concludes aims to collate succinctly the various data presented, making particular use of the Ashby plot approach. We also discuss the design criteria for different applications of NLC materials and summarise the various directions in which we expect the field to develop over the coming years.

\section{Compressibility: Theory and Measurement}

In the simplest terms, the compressibility of a material describes the relative rate of collapse of its linear dimensions with respect to pressure, measured or calculated at constant temperature:\cite{Nye:1984,Newnham:2005}
\begin{equation}\label{compress}
K_\ell=-\left(\frac{\partial\ell}{\ell\,\partial p}\right)_T.
\end{equation}
The minus sign means that positive compressibilities correspond to length reduction under increasing pressure. Conventional engineering materials such as steel and concrete contract by $\sim$0.5\% in every direction for each GPa of applied pressure, corresponding to a linear compressibility $K\sim5$\,TPa$^{-1}$.\cite{Ledbetter:1973}  Compressibility magnitudes usually reflect bond strengths, and so softer materials such as polymers and foams exhibit much larger values; for example, the linear compressibility of polystyrene is $K\simeq100$\,TPa$^{-1}$.\cite{Hellwege:1962}

Crystalline materials will in general have different compressibilities in different directions. For example, a layered material will usually be more compressible along the stacking axis than it is along a perpendicular direction.\cite{Munn:1972} This directional dependence can be relatively complex, especially when the crystal symmetry is low. We proceed to introduce the theory of compressibility in its most general form before explaining how the situation can be simplified as symmetry increases. Our starting point is the formal definition of compressibility as a rank-2 tensor:\cite{Nye:1984}
\begin{equation}\label{tensor}
\mathbf K=-\frac{\partial}{\partial p}\left[\begin{array}{lll}\epsilon_{11}&\epsilon_{12}&\epsilon_{13}\\ \epsilon_{21}&\epsilon_{22}&\epsilon_{23}\\ \epsilon_{31}&\epsilon_{32}&\epsilon_{33}\end{array}\right].
\end{equation}
Here the $\epsilon_{ij}$ are functions of hydrostatic pressure $p$ and represent the pressure-induced strain experienced by axis $j$ along axis $i$. The eigenvectors of Eq.~\eqref{tensor} describe an orthogonal coordinate system that brings $\mathbf K$ into diagonal form. These vectors are the so-called ``principal axes'' of compressibility (sometimes labelled $\mathbf x_1,\mathbf x_2,\mathbf x_3$) which can be interpreted as the crystal directions along which hydrostatic compression does not lead to any shear component. The eigenvalues of $\mathbf K$, which we term $K_1,K_2,K_3$, correspond to the compressibilities along these principal axes and are the unique descriptors of linear compressibility for any crystalline material. The formal requirement for NLC is that at least one of the $K_i$ is negative.

Defined in this way, the linear principal compressibilities are directly related to the volume compressibility, and in turn to the bulk modulus:
\begin{eqnarray}
K_V=-\left(\frac{\partial V}{V\,\partial p}\right)_T={\rm{Tr}}(\mathbf K)&=&K_1+K_2+K_3,\\
B=K_V^{-1}&=&\frac{1}{K_1+K_2+K_3}.
\end{eqnarray}
Because the volume compressibility must be positive, any system for which one of the linear compressibilities exceeds the bulk compressibility (\emph{i.e.}, $K_i>K_V=B^{-1}$) must exhibit NLC. This is the type of approach to identifying NLC materials employed in Ref.~\citenum{Baughman:1998}.

(A brief aside---Conventions vary in terms of the symbols used to denote these various elastic parameters. Compressibilities are denoted by some using the symbol $\beta$,\cite{Nye:1984} which is used by others to mean the volumetric coefficient of thermal expansion,\cite{Hook:1991} and by perhaps very many more to mean one of the unit cell angles. Likewise the bulk modulus is denoted by $K$ within much of the mineralogical literature, despite this symbol assuming the inverse meaning of compressibility when used in a physics text.\cite{Ashcroft:1976} In this review we adopt the conventions of the condensed matter physics community---\emph{i.e.}\ $K$ for compressibility and $B$ for bulk modulus---which we feel are the least likely to cause confusion.)

\subsection{Compressibilities from variable-pressure crystallographic measurements}

As mentioned above, the tensor algebra associated with compressibility determination is simplified enormously by consideration of crystal symmetry. For systems of orthorhombic crystal symmetry or higher, the principal axes coincide with the crystal axes. This means that the lattice parameter compressibilities 
\begin{eqnarray}
K_a&=&-\frac{1}{a}\left(\frac{\partial a}{\partial p}\right)_T,\\
K_b&=&-\frac{1}{b}\left(\frac{\partial b}{\partial p}\right)_T,\\
K_c&=&-\frac{1}{c}\left(\frac{\partial a}{\partial p}\right)_T,
\end{eqnarray}
which can be determined using variable-pressure crystallographic measurements, give directly the principal axis compressibilities. In other words, the $K_i$ reflect the relative rate of change of the lattice parameters with respect to pressure, and NLC materials can be identified as those for which at least one lattice parameter increases under hydrostatic pressure.

Unfortunately this equivalence between lattice and principal axis compressibilities does not hold for systems with monoclinic or triclinic crystal symmetries; lattice parameter compressibilities can have very little direct physical meaning in these cases.\cite{Cliffe:2012} In particular a negative value of one or more lattice parameter compressibilities would no longer imply NLC because the principal axis compressibilities may nonetheless remain positive. For such situations, there are software packages that facilitate the conversion from lattice parameter to principal axis compressibilities: PASCal and EoSfit are two examples.\cite{Cliffe:2012,Angel:2014}

\begin{figure}
\includegraphics{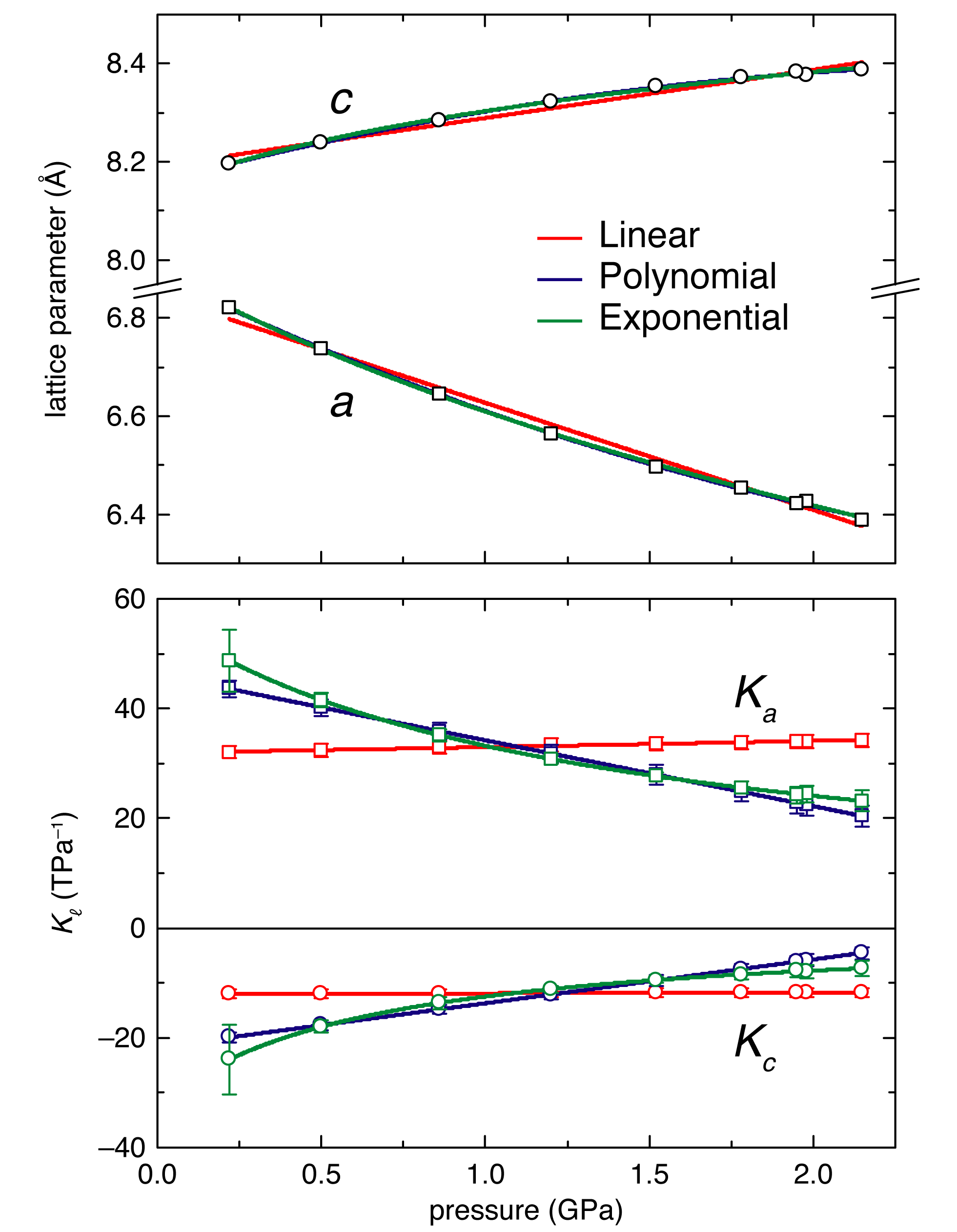}
\caption{Extraction of linear compressibilities from variable-pressure lattice parameter measurements. The top panel shows representative lattice parameter data for a smoothly-compressible material---in this case measured for KMn[Ag(CN)$_2$]$_3$.\cite{Cairns:2012} The three common approaches to fitting lattice parameter data represented by Eqs.~\eqref{lineareq}--\eqref{parameterisation} give the fits to data shown in red, blue, and green, respectively. The corresponding compressibilities are shown in the lower panel. All three methods obtain comparable average compressibility values over the entire pressure range for which data are fitted. The reduction in magnitude of $K$ at higher pressures is unaccounted for by the linear fitting method, and is treated slightly differently by the two non-linear fits. The extrapolated values of $K$ at $p=0$ are very different in all three cases.}
\label{fig2}
\end{figure}

One further complication in converting variable-pressure lattice parameters to linear compressibilities is the tendency for lattice parameters to depend non-linearly on pressure; \emph{i.e.} the $K_i$ are themselves pressure-dependent. (Indeed if any $K_i$ were truly constant then there would exist a finite pressure at which the corresponding material length would vanish: $p_{\rm{crit}}=1/K_i$). There is no thermodynamic requirement for the $K_i$ to depend on pressure in any particular way; this situation contrasts that of the pressure-dependence of the crystal volume, which is often interpreted in terms of the Birch-Murnaghan equations of state.\cite{Birch:1947,Sata:2002} Instead the lattice parameters are usually fitted to some empirical parameterisation of choice. In the simplest case this would be the linear relationship
\begin{equation}\label{lineareq}
\ell(p)=\ell_0[1-K_\ell p],
\end{equation}
where $\ell_0$ represents the length at zero pressure. The nonlinearity left unaccounted for by this simple parameterisation can be included \emph{via} higher-order polynomial expansions:
\begin{equation}
\ell(p)=\ell_0\left[1+\sum_{i=1}^n\alpha_ip^i\right].
\end{equation}
Some authors then identify the value $-\alpha_1$ with the linear compressibility $K_\ell$; however, this tends to overestimate the compressibility in cases where there is a strong pressure dependence.\cite{Goodwin:2008a,McCann:1972,Mariathasan:1985} In our own work, we have found that the alternative parameterisation
\begin{equation}\label{parameterisation}
\ell(p)=\ell_0+\lambda(p-p_{\rm c})^\nu
\end{equation}
actually captures better the pressure-dependence of lattice parameters for most systems. The corresponding compressibilities are determined straightforwardly from the pressure derivative of Eq.~\eqref{parameterisation}:
\begin{equation}
K(p)=-\frac{1}{\ell(p)}\lambda\nu(p-p_{\rm c})^{\nu-1}.
\end{equation}
The value of $K(p)$ determined in this way diverges at $p=p_{\rm c}$, since $\nu<1$; often this divergence has physical significance in terms of the elastic instability at a pressure-induced phase transition. Figure~\ref{fig2} compares the linear compressibilities determined in these different ways for some representative variable-pressure lattice parameter data.

Whichever parameterisation is used, accurate determination of the $K_i$ generally relies on access to a relatively large number of lattice parameter measurements over the pressure interval of interest. Experimental claims of NLC are sometimes made on the basis of just two measurements;\cite{Paliwoda:2014} however a general rule of thumb is that 10 measurements are needed for accurate compressibility determination with little improvement for more than 20 measurements.\cite{Hazen:2000}

\subsection{Compressibilities from elastic compliances}

Because lattice compressibilities are one aspect of the more general elastic behaviour of materials, determination of the elastic stiffness tensor $\mathbf C$ provides an alternative means of characterising NLC.\cite{Newnham:2005} The relationship between the $K_i$ and $\mathbf C$ is most straightforwardly established by considering the elements of the elastic compliance tensor $\mathbf S$ (the inverse of $\mathbf C$), which relate strains $\epsilon_{ij}$ to applied stresses $\sigma_{ij}$:
\begin{equation}\label{bigseqn}
\epsilon_{ij}=-\sum_{k,l=1}^3S_{ijkl}\sigma_{kl}.
\end{equation}
Here all terms are defined in the limit of infinitesimal strain. Eq.~\eqref{bigseqn} is essentially a form of Hooke's law that relates displacement ($\propto$ strain) to mechanical force ($\propto$ stress). Hydrostatic compression is the specific situation where axial stresses are all equal to the applied pressure and shear stresses are absent; that is,
\begin{eqnarray}
\sigma_{ii}=p& &\forall i,\\
\sigma_{ij}=0 & &\forall i\neq j.
\end{eqnarray}
By design, the principal axes are those for which axial compression does not induce any shear strain, and so Eq.~\eqref{bigseqn} reduces to
\begin{equation}
\epsilon_{ii}=-p\sum_{k=1}^3S_{ik}
\end{equation}
when expressed in the principal axis coordinate system (noting that we have now switched to Voigt notation so that $S_{ik}\equiv S_{iikk}$). Substitution into Eq.~\eqref{tensor} gives the simple relationship
\begin{equation}\label{kicalc}
K_i=\sum_{j=1}^3S_{ij}.
\end{equation}
that often appears in texts on the subject.\cite{Newnham:2005} Consequently, determination of the elastic stiffness tensor $\mathbf C$ (either experimentally or computationally) can also yield the linear compressibilities $K_i$ via the compliance tensor $\mathbf S=\mathbf C^{-1}$.

Experimental techniques capable of probing the tensor $\mathbf C$ include resonant ultrasound spectroscopy (RUS),\cite{Migliori:1997} Brillouin scattering,\cite{Sanchez-Valle:2005,Tan:2012} inelastic neutron scattering,\cite{Shirane:2002} nanoindentation\cite{Tan:2010} and shear-wave velocity\cite{Peercy:1975} measurements. In most cases these measurements are indirectly sensitive to a subset of the elements, or combinations of elements, of $\mathbf C$. Consequently interpretation of the experimental data is often carried out \emph{via} a parameterised lattice-dynamical model (containing fewer free parameters than $\mathbf C$ itself), from which $\mathbf C$ is subsequently calculable using software packages such as GULP.\cite{Gale:1997} The single most important distinction between experimental compressibilities determined \emph{via} elastic compliances and those obtained from variable-pressure crystallographic measurements is that the former correspond to values obtained \emph{in the limit of zero applied pressure} and hence are usually much larger in magnitude.

\emph{Ab initio} methods also allow determination of uniaxial compressibilities in an analogous way. Starting from the optimised geometry for a given crystal, sampling of every possible pairwise combination of strains $\epsilon_{ij},\epsilon_{kl}$ allows the $C_{ijkl}$ to be evaluated directly from the corresponding second derivatives of the lattice energy:\cite{Coudert:2013}
\begin{equation}
C_{ijkl}=\frac{1}{V}\left(\frac{\partial^2E}{\partial \epsilon_{ij}\partial \epsilon_{kl}}\right).
\end{equation}
The tensor $\mathbf S$ is obtained by inversion and the $K_i$ calculated according to Eq.~\eqref{kicalc}. Such an approach is implemented in, for example, the CRYSTAL09 code,\cite{Dovesi:2005,Perger:2009} and has been applied to the exploration of negative compressibility of simple inorganics,\cite{Barnes:2011,Marmier:2010} zeolites,\cite{Coudert:2013} and metal--organic frameworks alike.\cite{Ortiz:2012,Ortiz:2013} Once again, the compressibilities obtained in this way represent the zero-pressure limit and as such can be vastly more extreme than those determined across finite pressure ranges.\cite{Ortiz:2012}


This link between linear compressibilities and the more general elastic properties of materials means that the observation of NLC is often diagnostic of other anomalous mechanical responses. Examples include extreme mechanical anisotropy,\cite{Ortiz:2012} unusual values of the Poisson's ratio,\cite{Goodwin:2008a} negative thermal expansion (NTE),\cite{Munn:1972,Ogborn:2012} and a propensity towards dynamic instabilities.\cite{Goodwin:2008a} Importantly, this correspondence works both ways. Severe structural anisotropy (such as arising from the preferred orientations of molecules in a particular packing arrangement or the symmetry of a given framework topology), or the observation of uniaxial or biaxial NTE are increasingly frequently found to be strong predictors of NLC behaviour.\cite{Cairns:2013a,Cai:2014} We will come to explore this correspondence in more detail below.

\subsection{Compressibility capacity}

Because of the pressure dependence of the $K_i$ it is important that compressibility values are quoted with reference to the pressure range over which they have been determined. This pressure range may correspond to the entire stability field of the phase in question, may be imposed by experimental constraints, or may be of relevance to a specific industrial process (by way of example, most machining processes subject materials to pressures of \emph{ca} 1--2\,GPa\cite{Chapman:2009}). One obvious limitation in comparing the NLC behaviour of different materials in terms of the magnitudes of $K_i$ alone is that these values may be determined over very different pressure ranges for different materials.

A metric that we have found useful in comparing the degree of NLC behaviour for different materials is what we term the compressibility capacity:
\begin{equation}
\chi_K=-\int_0^{p_{\rm c}}K(p)\,{\rm d}p,
\end{equation}
where $p_{\rm c}$ denotes the maximum pressure for which NLC is observed. The value of $\chi_K$ is a dimensionless quantity that takes into account both the magnitude of NLC and the pressure range over which NLC occurs. It simply represents the maximum total fraction by which a crystal can expand under application of hydrostatic pressure. We will report $\chi_K$ values for the various NLC materials reviewed in this article as and when each is introduced; anticipating these results, we find that $\chi_K\sim$1\% for the majority of known NLC systems, with values closer to 10\% only observed in the most exceptional cases.

\section{NLC Materials}

What follows is an overview of the various materials known to exhibit NLC. We have grouped these into four classes according to the microscopic mechanism likely to be responsible for NLC in each case: (i) those compounds for which NLC arises as a consequence of proper, improper, or quasi-ferroelastic phase transitions; (ii) network solids for which NLC is driven by correlated polyhedral tilts; (iii) helical systems; and (iv) framework materials with wine-rack, honeycomb, or related topologies, where NLC arises from framework hinging. In this way, we hope to summarise not only the properties of known NLC materials but also the mechanisms that can give rise to the phenomenon itself.


\subsection{Ferroelastics}

\begin{figure}[b]
\begin{center}
\includegraphics{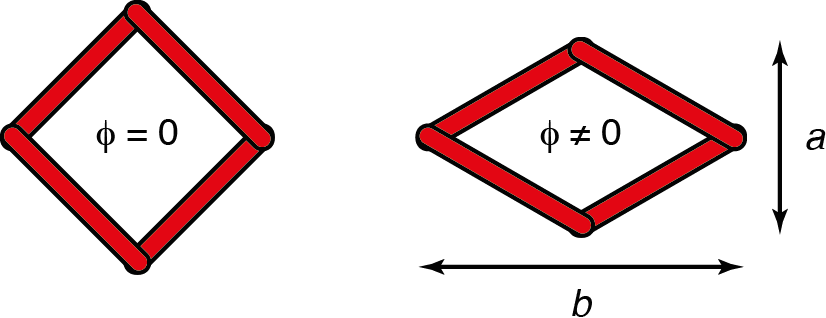}
\caption{The paraelastic (left) and ferroelastic (right) states of the square lattice. The ferroelastic order parameter $\phi=(b-a)/(a+b)$ is a measure of the extent of symmetry-breaking distortion.}
\label{fig3}
\end{center}
\end{figure}

Ferroelasticity describes the emergence of spontaneous strain in a symmetry-breaking phase transition.\cite{Salje:1993} The phenomenon can be considered the mechanical equivalent of ferroelectricity or ferromagnetism, where it is the spontaneous strain---rather than polarisation or magnetisation---that behaves as the ferroic order parameter. A simple example is the square $\rightarrow$ rhombic transition illustrated in Fig.~\ref{fig3}, for which the ferroelastic order parameter is a measure of the distortion away from square symmetry:\cite{Salje:1993a,Redfern:1988,Haines:1997}
\begin{equation}\label{spontstrain}
\phi=\frac{b-a}{a+b}.
\end{equation}
The value of $\phi$ is zero in the `paraelastic' high-symmetry phase and non-zero in the ferroelastic low-symmetry phase (square and rhombic geometries in Fig.~\ref{fig3}, respectively). One result of Landau theory is that any transition for which strain is the primary order parameter must be second-order in nature.\cite{Salje:1993a} So, there being no volume discontinuity across these so-called `proper ferroelastic transitions', the emergence of spontaneous strain (\emph{i.e.}\ $\phi\neq0$ or, equivalently, $a\neq b$ in Eq.~\eqref{spontstrain}) requires at least one axis to increase in length on symmetry lowering. This means that we can expect NLC in any system that supports a pressure-induced proper ferroelastic transition: the lower-symmetry phase stabilised at pressures above the transition must expand along at least one crystallographic axis for at least some finite pressure interval.

\subsection*{Rutiles}

\begin{figure}[t]
\begin{center}
\includegraphics{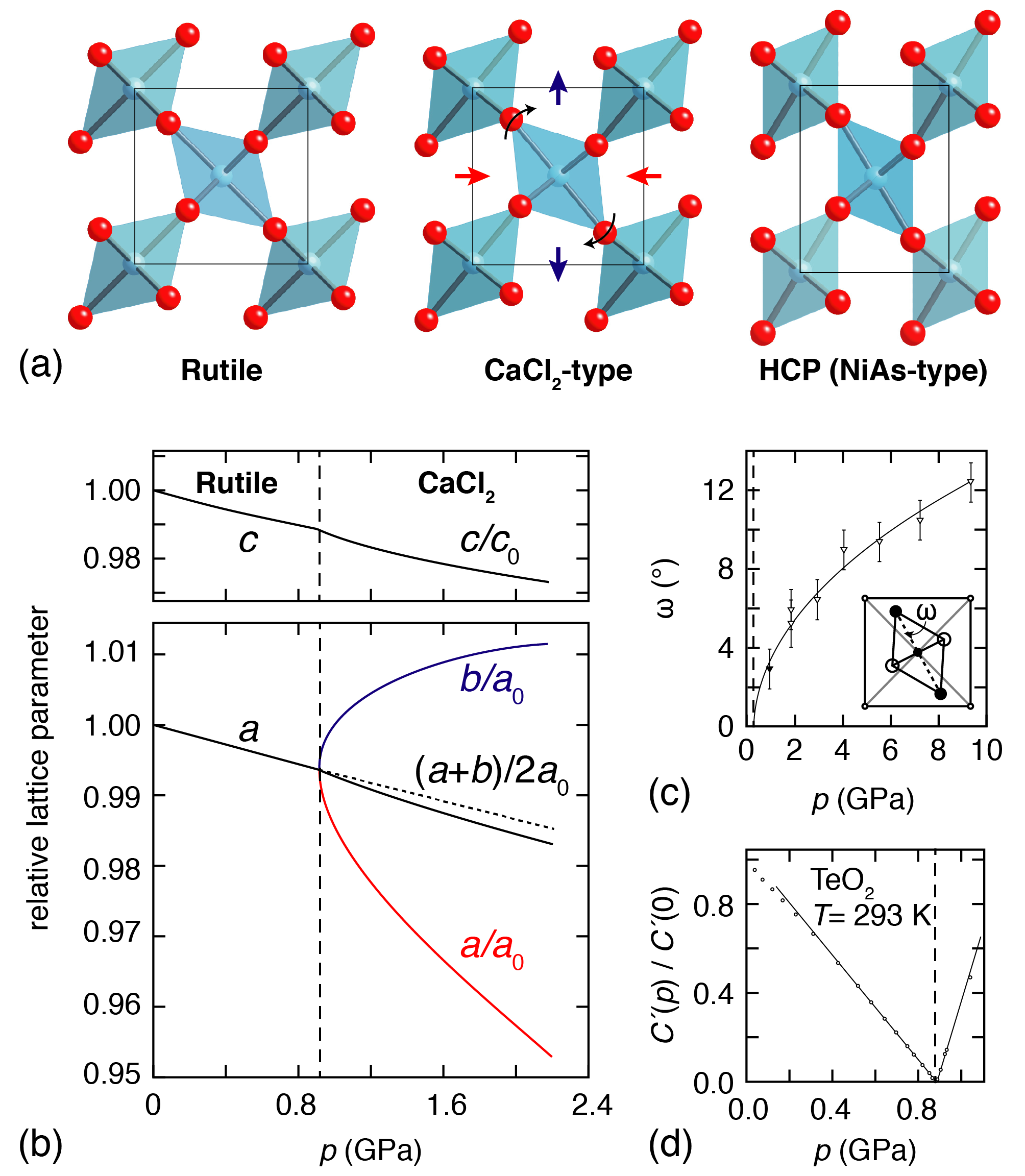}
\caption{Ferroelastic NLC mechanism in rutiles. (a) The ferroelastic instability of the rutile structure type corresponds to a progressive distortion towards a defect NiAs structure and involves expansion in the vertical direction.\cite{Ming:1980} (b) Pressure-dependence of the lattice parameters of TeO$_2$, showing the ferroelastic transition at $\sim0.81$\,GPa.\cite{Worlton:1975} The high-pressure CaCl$_2$-structured phase exhibits NLC along $\mathbf b$. (c) Evolution of the octahedral tilt parameter $\omega$.\cite{Haines:1995} (d) Softening of the effective elastic constant $C^\prime$ at the ferroelastic transition.\cite{Peercy:1975}}
\label{fig4}
\end{center}
\end{figure}

\begin{table*}
\small
  \caption{\ Compressibilities of materials for which NLC arises as a consequence of ferroelastic or related phase transitions.}
  \label{tbl:rutiles}
  \begin{tabular*}{\textwidth}{@{\extracolsep{\fill}}llllllllll}
    \hline
	 			& Method		&$ K_1$ (TPa$^{-1}$)	& $K_2$ (TPa$^{-1}$)	& $K_3$ (TPa$^{-1}$)	& $B_0$ (GPa)$^{a}$ 	 	& Range (GPa) 	&$\chi_{_{K}}$ (\%)		& Ref(s). \\ \hline
TeO$_2$ 			& $^a$		&$-$5.1(6)			&2.1(7)				& 18.4(6) 				& 52(4) 				 	& 0.9--3.25 		&1.20(14)				& \citenum{Skelton:1976, McWhan:1975, Peercy:1975, Worlton:1975,Uwe:1979}\\
NiF$_2$			& $^a$ 		&$-$0.48				&0.61				&4.41				& 222 				 		& 1.8--3.2 			& 0.067				& \citenum{Jorgensen:1978, Mirgorodsky:1994}\\
$\beta$-MnO$_2$ 	& $^a$ 		& {$-$0.16(7)}			& 0.269(17) 			& 1.82(10)   			& 328(18) 						& 0.3--29.3 		& 0.46(20)				& \citenum{Haines:1995} \\
MgF$_2$			& $^a$ 		&{$-$1.3(3)} 			& 2.51(2) 				& 8.05(10) 			& 68(13) 				 	& 9.1--10.4 		& 0.17(4)				& \citenum{Haines:2001, Zhang:2008} \\
PbO$_2$-I$^{\prime}$&$^a$ 		& {$-$1.82(15)}			& 1.76(7) 				& 3.40(3)    			& 167(18) 				 	& 3.8--6.1 			& 0.42(3)				& \citenum{Haines:1996} \\
GeO$_2$			&$^a$ 		&{$-$0.137(12)}		& 0.827(14) 			& 2.02(13)  			& --- 							& 28--36 			& 0.110(10)			& \citenum{Haines:2000} \\
SnO$_2$			&$^a$		&{$-$0.185} 			& 0.394				&  2.345  				& 204(6) 						&11.8--21 			& 0.170				& \citenum{Haines:1997, Endo:1990}\\    \hline
Zn(CN)$_2$-II			&$^b$	& $-$2.08 				& 11.2 				& 16.8 				&  ---							&1.52--5 			& 0.724				& \citenum{Collings:2013, Lapidus:2013} \\
Pb$_{3}$(PO$_{4}$)$_2$	&$^a$ 	& {$-$4.3(4)}			&5.0(2)  				& 20.98(18)  			& 38.7(5) 				 	&0--1.59 			&0.68(6)				&\citenum{Angel:1999,Angel:2004a,Salje:1981}  \\ 
InS					&$^{a}$ 	 & {$-$2.41(13)}		& 2.8(4) 				& 15.3(22) 			& 33.2(18) 			  	& 0-4.3 			&1.04(6)				& \citenum{Kabalkina:1982, Schwarz:1995, Takarabe:1988} \\ \hline
Sillimanite				&$^{a,c}$ 	& {$-$3.30} 			& 1.45 				&10.8 				& 112.50 						& 29.9--37.5 		&2.51				& \citenum{Oganov:2001}\\
PtS					&$^{b,c}$ 	& {$-$0.47 -- $-$0.92}	& 3.3 				& 3.3  				& 167.5		 			& 0--10 			&0.47--0.92			& \citenum{Marmier:2010, Collins:1979}\\ \hline
  \end{tabular*}
  $^{a}$ Calculated from PASCal\cite{Cliffe:2012} from reported variable-pressure lattice parameters. $^b$ As reported. $^{c}$ From DFT calculation.
\end{table*}

A number of simple binary inorganic solids with the TiO$_2$ rutile structure exhibit NLC \emph{via} precisely this mechanism.\cite{Ming:1980} The well-known tetragonal crystal structure of rutile consists of columns of edge-sharing octahedra which are in turn connected at their corners [Fig.~\ref{fig4}(a)]. This structure has a ferroelastic instability associated with correlated rotations of neighbouring columns of octahedra.\cite{Peercy:1975,Haines:1995} Activation of this tilt system lowers the crystal symmetry from tetragonal to orthorhombic, resulting in the (equally well-known) CaCl$_2$ structure [Fig.~\ref{fig4}(a)]. In order to conserve volume, the lattice is forced to expand along one of the two directions perpendicular to the column axis.\cite{Anderson:1965,Worlton:1975} This transition can be viewed as a progression towards the defect-NiAs-type structure of FeS$_2$ marcasite [Fig.~\ref{fig4}(a)] and is thought to be driven largely by considerations of anion packing efficiency.\cite{Ming:1980}

A representative example of a pressure-induced ferroelastic transition---and hence NLC---arising from this type of instability is given by the mineral paratellurite (TeO$_2$).\cite{Skelton:1976, McWhan:1975, Worlton:1975,Uwe:1979} Variable-pressure lattice parameter measurements for this material reveal a ferroelastic transition at a critical pressure $p_{\rm c}\simeq0.9$\,GPa, followed by NLC along the $b$ crystal axis of the resulting high-pressure phase [Fig.~\ref{fig4}(b)]. The proposed mechanism of correlated octahedral tilts can be validated by determining the pressure-dependence of the octahedral tilt angle $\omega$, which also behaves as an order parameter for the transition [Fig.~\ref{fig4}(c)].\cite{Haines:1995} That the transition is truly strain-driven can be deduced from the elastic behaviour near $p_{\rm c}$: the effective elastic constant $C^\prime=\frac{1}{2}(C_{11}-C_{12})$ governs the relevant shear mode velocity, and can be seen to vanish at the transition point [Fig.~\ref{fig4}(d)].\cite{Peercy:1975,Toledano:1983}

\begin{figure}[t]
\begin{center}
\includegraphics{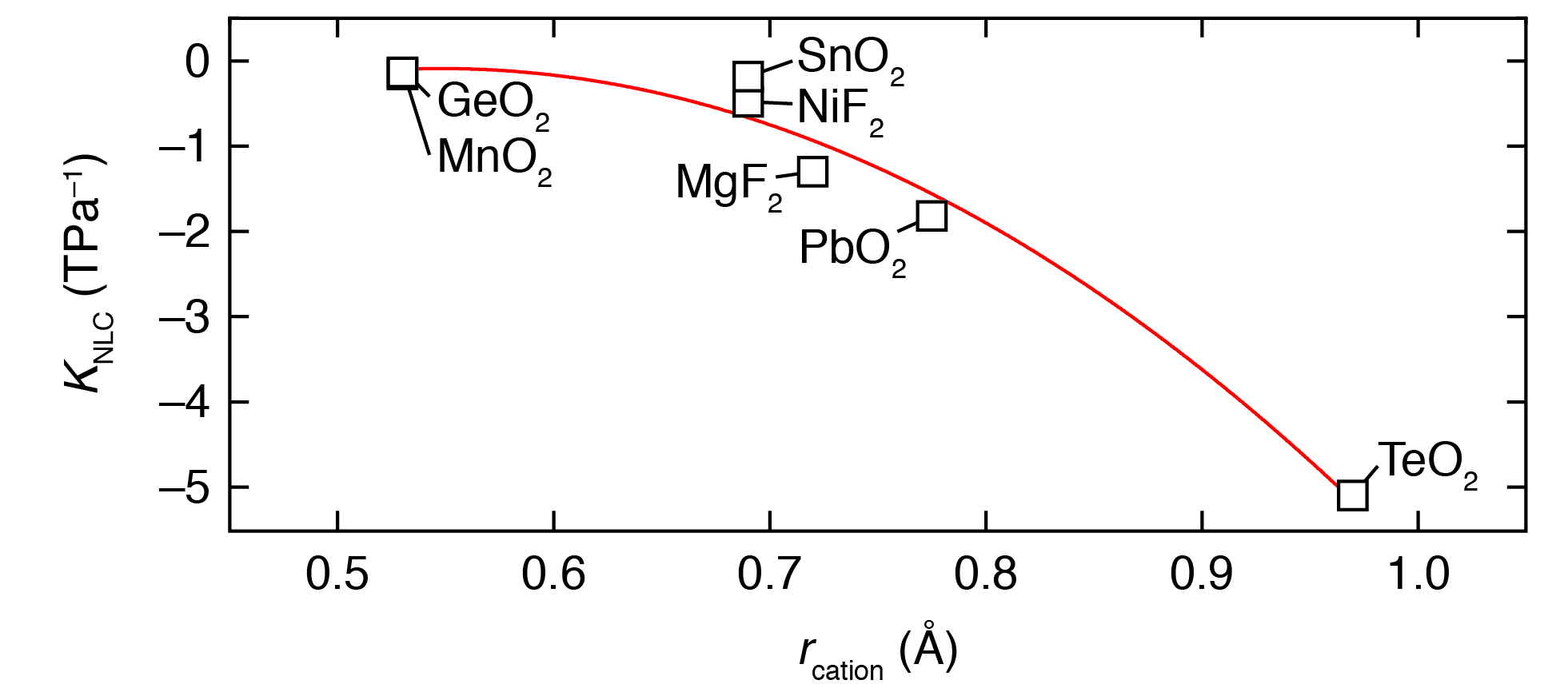}
\caption{Relationship between magnitude of NLC and cation radius for rutile-structured dioxides and difluorides. Radii taken from Ref.~\citenum{Shannon:1976}. The line is a guide to the eye.}
\label{fig4e}
\end{center}
\end{figure}

TeO$_2$ is not an isolated example of this behaviour: a number of rutile-structured dioxides and difluorides exhibit NLC \emph{via} essentially the same mechanism.\cite{Ming:1980} The magnitude of NLC for these different materials can be determined from the various crystallographic measurements reported in the literature. In the case of TeO$_2$, the data of Ref.~\citenum{Uwe:1979}\ give $K_{\rm{NLC}}=-5.1(6)$\,TPa$^{-1}$ over the pressure range 0.9--3.25 GPa, corresponding to a compressibility capacity of $\chi_K=1.20(14)$\%. Table~\ref{tbl:rutiles} compares these values for the various rutiles for which a high-pressure ferroelastic phase transition has been observed. What emerges is that there is a general correspondence between cation radius and extent of NLC, such that TeO$_2$---which contains the largest\cite{Shannon:1976} of the cations---also shows the most extreme NLC response [Fig.~\ref{fig4e}]. We will show by comparison with other families that a compressibility capacity $\chi_K\simeq1$\% is not itself particularly extreme . So while it is the case that a ferroelastic instability mechanism for NLC may give rise to relatively general behaviour, it would seem that---at least in the case of rutiles---the mechanism is unlikely to give rise to especially large NLC responses. One additional complication is that the NLC effect is of course only observed in the non-ambient phase, so future research within this family might likely concentrate on lowering $p_{\rm c}$ (perhaps even to negative pressures; \emph{i.e.}, studying CaCl$_2$-structured materials) and/or varying cation/anion radii so as to allow the largest possible spontaneous strains to emerge within the ferroelastic phase.

\subsection*{Zinc cyanide: an improper ferroelastic}

Even materials for which hydrostatic pressure induces a so-called `improper' ferroelastic phase transition can exhibit NLC within the high-pressure phase. The label `impoper' simply means that the spontaneous strain is no longer the primary order parameter responsible for driving the phase transition.\cite{Salje:1993} Instead ferroelastic strain develops \emph{via} coupling to an alternate, dominant, symmetry-breaking mechanism, such as a phonon instability.\cite{Fradkin:1997} The Landau conditions change in these situations such that volume discontinuities may be observed, and so there is no strict guarantee of NLC. Nevertheless if the volume collapse on symmetry lowering is small with respect to the subsequent evolution of ferroelastic spontaneous strain, then NLC may indeed arise.

\begin{figure}
\begin{center}
\includegraphics{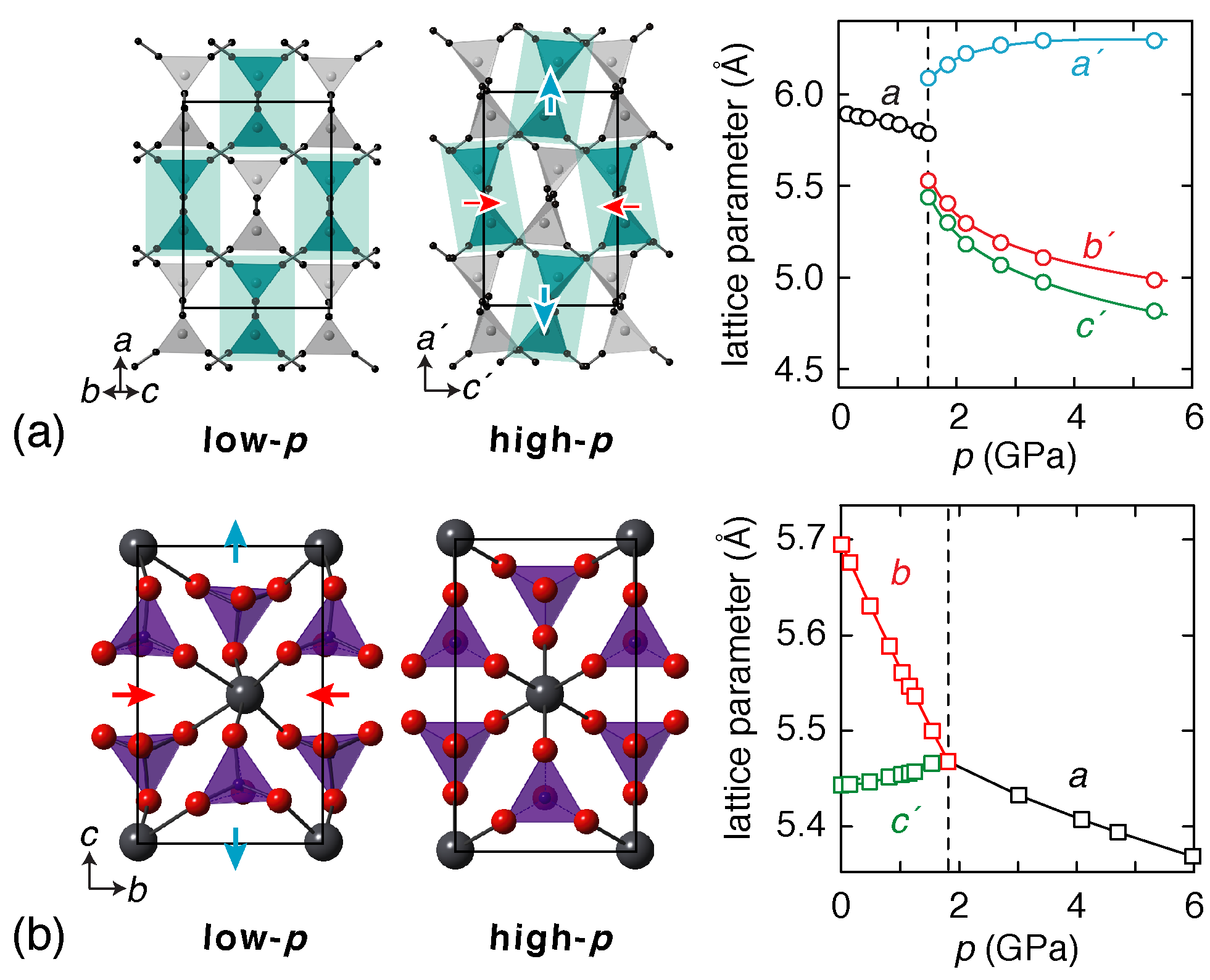}
\caption{NLC from improper and reverse ferroelastic transitions. (a) The cubic $\rightarrow$ orthorhombic improper ferroelastic transition in Zn(CN)$_2$ involves correlated rotations of column pairs of Zn(C/N)$_4$ tetrahedra (shaded in green), resulting in NLC along the $\mathbf a$ axis of the daughter cell.\cite{Collings:2013} (b) The ferroelastic state of Pb$_3$(PO$_4$)$_2$ is denser than the paraelastic parent, and so application of pressure induces a reverse ferroelastic transition. The $\mathbf c$ parameter of the ambient phase (the length of which is normalised here $c^\prime=c/\sqrt{3}$ for comparison) expands as the transition is approached.\cite{Angel:2004a} Pb atoms are shown as large black spheres and PO$_4$ units as filled tetrahedra.}
\label{fig5}
\end{center}
\end{figure}

This is precisely the mechanism that appears to be responsible for NLC in the high-pressure phase of the molecular framework material zinc cyanide, Zn(CN)$_2$.\cite{Collings:2013,Fang:2013,Lapidus:2013} Under ambient conditions Zn(CN)$_2$ adopts a cubic structure in which Zn$^{2+}$ cations are tetrahedrally coordinated by four cyanide anions, each of which in turn connects two Zn centres to give a three-dinemensional framework with the diamond-like anti-cuprite topology.\cite{Williams:1997} Both computational studies and inelastic neutron scattering measurements point to the existence a large family of soft phonon modes, many of which are also implicated in the strong NTE behaviour observed experimentally.\cite{Fang:2013,Goodwin:2005,Chapman:2006a} Hydrostatic compression to 1.52\,GPa results in a first-order displacive phase transition to a denser orthorhombic structure, which is consistent with the condensation of at least one zone-boundary soft mode.\cite{Collings:2013,Lapidus:2013} The corresponding atomic displacements involve correlated rotations of connected pairs of Zn(C/N)$_4$ tetrahedra, resulting in coupled expansion/contraction of the crystal lattice perpendicular to the rotation axis in much the same way as described for the rutile-structured systems discussed above [Fig.~\ref{fig5}(a)].\cite{Collings:2013} NLC persists in the high-pressure Zn(CN)$_2$-II phase from the I/II transition at 1.52\,GPa up to $\sim$5\,GPa and is reasonably strong over this entire pressure range: $K_{\rm{NLC}}=-11(3)$\,TPa$^{-1}$ and $\chi_K=3.8(10)$\%.

In competition with any ferroelastic NLC mechanism is the tendency for framework buckling, which favours PLC and becomes increasingly important at higher pressures.\cite{Ogborn:2012} In the specific case of Zn(CN)$_2$-II this buckling involves a systematic distortion of Zn(C/N)$_4$ polyhedra and bending of Zn--C--N--Zn linkages to allow additional neighbouring cyanide ions within the originally-tetrahedral Zn coordination sphere.\cite{Collings:2013} The Zn coordination number progressively increases from four to six, and the resulting volume reduction becomes increasingly significant with respect to the NLC effect of correlated tilts, such that the crystal axis along which NLC is initially observed begins to contract for pressures higher than 5\,GPa. Such crossover between NLC and PLC behaviour might be expected to be a reasonably general phenomenon, and has certainly be noticed previously in \emph{e.g.}\ a variable-pressure study of the metal--organic framework silver(I) methylimidazolate.\cite{Ogborn:2012}

\subsection*{Reverse ferroelastics: lead phosphate and indium sulfide}

Lead phosphate, Pb$_3$(PO$_4$)$_2$, is the unusual example of a material that exhibits a `reverse' ferroelastic phase transition on compression: the low-symmetry monoclinic structure converts to a higher-symmetry, denser, rhombohedral phase at a hydrostatic pressure of approximately 1.6\,GPa.\cite{Angel:2004,Wood:1980} Because ferroelastic transitions couple expansion and contraction along orthogonal axes, NLC is actually expected irrespective of the direction in which the transition between high symmetry and low symmetry states is traversed. Indeed Pb$_3$(PO$_4$)$_2$ shows NLC throughout the entire stability field of the ferroelastic phase with $K_{\rm NLC}=-4.3(4)$\,TPa$^{-1}$ and $\chi_K=0.68(6)$\%.\cite{Angel:2004a,Decker:1979} The mechanism responsible for NLC---which resembles that observed on heating Pb$_3$(PO$_4$)$_2$ at ambient pressure---again involves correlated polyhedral tilts (in this of the PO$_4$ tetrahedra) which couple to off-centering of the Pb$^{2+}$ cations [Fig.~\ref{fig5}(b)]. Doping with Ba$^{2+}$ predictably favours the rhombohedral (cation-centred) state, lowering at once both transition pressure and temperature while preserving NLC.\cite{Decker:1979,Angel:2004a, Angel:1999,Salje:1981,Salje:1993}

A less-well characterised example of a possible reverse ferroelastic transition is that of indium(II) sulfide, InS [$\equiv$ (In$_2$)$^{4+}$(S$^{2-}$)$_2$].\cite{Takarabe:1988,Schwarz:1995} Certainly its ambient phase (orthorhombic $Pmnn$ symmetry) shows NLC over the pressure range 0--4.3\,GPa ($K_{\rm{NLC}}=-2.41$(13)\,TPa$^{-1}$), and this behaviour is qualitatively understandable in terms of the straightening of S--In--In--S ``dumbells''. The confusion lies in the relationship of this response to the high-pressure InS-II phase that forms at 7.5\,GPa. This high-pressure phase was originally reported to have the tetragonal Hg$_2$Cl$_2$ structure (\emph{i.e.} with linear S--In--In--S units), which is of exactly the right symmetry to be considered the paraelectric parent of the ambient phase.\cite{Kabalkina:1982} However, a more recent \emph{in situ} single crystal structure determination reports InS-II to adopt a distorted monoclinic structure in which the same S--In--In--S units remain buckled.\cite{Schwarz:1995} So while there is no ambiguity regarding the NLC effect itself, and there is an implied relationship to the existence and nature of a high-pressure phase transition, a detailed mechanistic understanding of this relationship would demand further experimental characterisation of the pressure-dependent behaviour of this system.

\subsection*{NLC from ferroelastic-like phase transitions}

\begin{figure}
\includegraphics{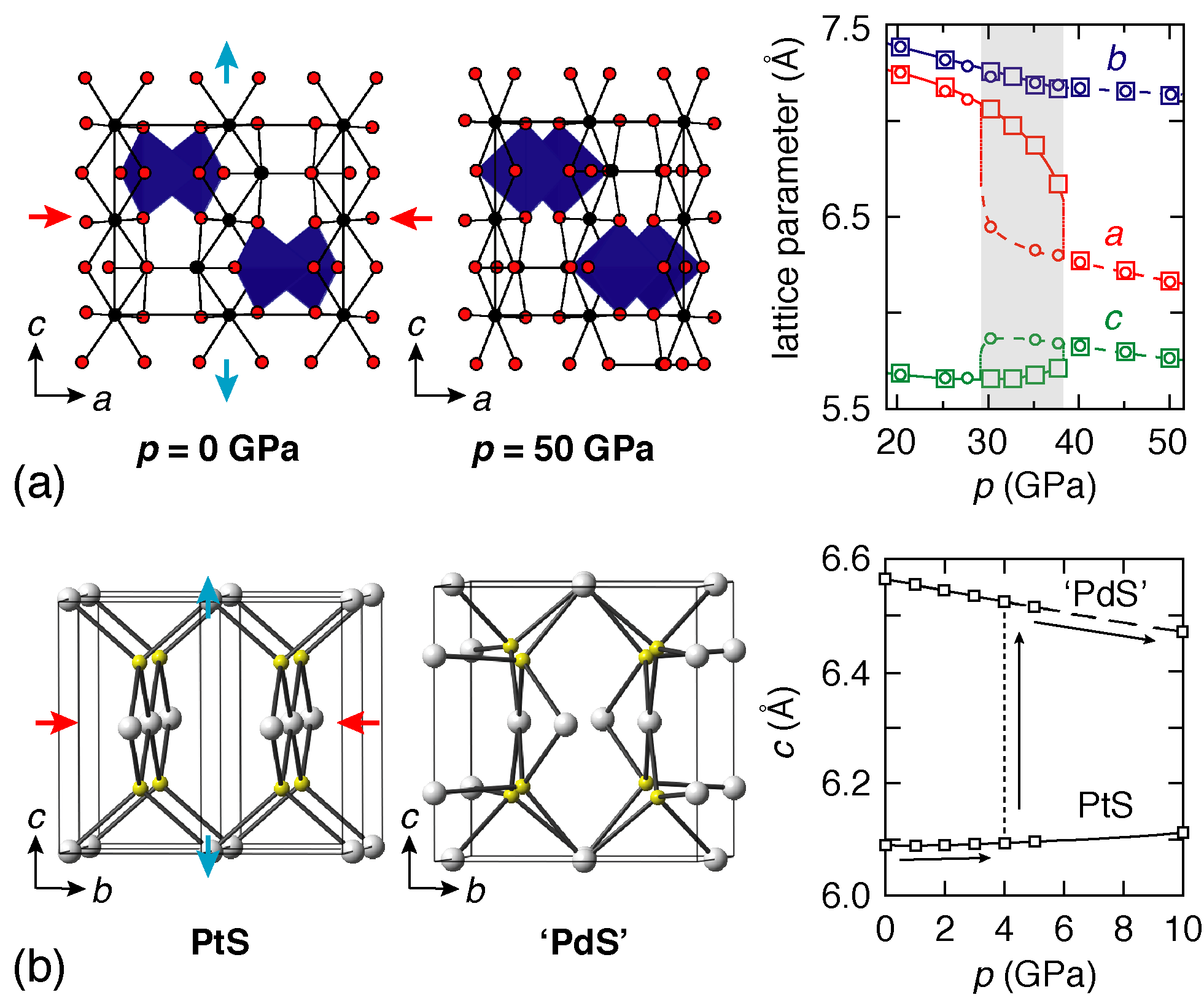}
\caption{(a) The first-order isosymmetric transition in sillimanite involves a discontinuous increase in Si coordination number from four to five (SiO$_n$ polyhedra shown in blue). The corresponding lattice rearrangement involves an expansion along the $\mathbf c$ axis.\cite{Oganov:2001} (b) At hydrostatic pressures of $\sim$3--4\,GPa, PtS undergoes a reconstructive transition to the denser PdS structure type.\cite{Marmier:2010,Collins:1979} Both phases have tetragonal symmetry but the latter expanded along $\mathbf c$ relative to the former. Quantum mechanical calculations show that, as the transition is approached on increasing pressure, the ambient phase is expected to expand along the same axis.\cite{Marmier:2010}}
\label{fig6}
\end{figure}

Other types of phase transition---which are not strictly ferroelastic, but which share mechanistic similarities---may also give rise to NLC. We illustrate this point with two final examples. The first involves a geologically-relevant isosymmetric transition in the sillimanite polymorph of Al$_2$SiO$_5$.\cite{Oganov:2001} The ambient phase of this framework structure contains four-coordinate Si atoms.\cite{Winter:1979} Lattice dynamical calculations using well-established interaction potentials optimised for aluminosilicates\cite{Winkler:1990} indicate that the framework undergoes a correlated reorganisation between 30 and 40\,GPa that increases the Si coordination number from four to five.\cite{Oganov:2001} All original bonding connectivity is preserved in this process (only new bonds are formed), and there is no change in crystal symmetry; such a transition is necessarily first-order in nature.\cite{Christy:1995,Bruce:1981} As a result of the increased Si coordination number, the high-pressure phase is denser than the low-pressure phase, but the new connectivity results in an increase in length along the $c$ axis of its orthorhombic $Pnma$ cell. As the transition is approached with increasing hydrostatic pressure, the gradual conversion of one phase to the other results in an NLC effect [Fig.~\ref{fig6}(a)]. 

Our second example is the transformation of PtS from its ambient-pressure structure (tetragonal $P4_2/nmm$) to a high-pressure phase with the so-called PdS structure (tetragonal $P4_2m$) at pressures of between 2.5 and 3.0\,GPa.\cite{Collins:1979} From an experimental viewpoint, very little is understood regarding this transition. What is known is that the high-pressure phase is metastable under ambient conditions and is expanded by $\sim$8\% along the tetragonal axis and compressed $\sim$9\% along the two perpendicular axes relative to the thermodynamic phase.\cite{Collins:1979} First principles calculations suggest that the deformation mechanism of the ambient phase under compression resembles a progression towards this high-pressure structure, which means that coupled PLC/NLC is expected, with NLC occurring along the tetragonal axis [Fig.~\ref{fig6}(b)].\cite{Marmier:2010} The magnitude of NLC obtained in these calculations turns out to depend on the particular functionals used: the authors of Ref.~\citenum{Marmier:2010} find $K_{\rm{NLC}}=-0.47$\,TPa$^{-1}$ for LDA and $-0.92$\,TPa$^{-1}$ for GGA (calculated over the pressure range 0--10\,GPa in both instances). The large expansion between the two phases observed experimentally (which places an upper limit on the true value of $\chi_K$) suggests that the real value of $K_{\rm{NLC}}$ may be higher than these calculated values, and clearly additional experimental characterisation would play a valuable role in understanding better the intriguing NLC behaviour of this system.

So the structural changes that occur near phase transitions can give rise to NLC for a variety of different structural families and different transition mechanisms. NLC is all but guaranteed in the case of proper ferroelastic transitions, given that the development of spontaneous strain without a volume discontinuity requires expansion of the crystal lattice along at least one direction. Ambient-pressure variable-temperature studies may provide a useful method of identifying likely NLC candidates for further study, by virtue of the empirical observation that ferroelastic instabilities observed on cooling are also often observed under hydrostatic pressure. In this respect the recent discovery that some ferroelastic metal--organic frameworks (MOFs) can develop extremely large spontaneous strains on cooling suggests that equally extreme NLC may also be discovered in the very same systems.\cite{Hunt:2015}

\subsection{Tilting networks}

\begin{figure}[b]
\begin{center}
\includegraphics{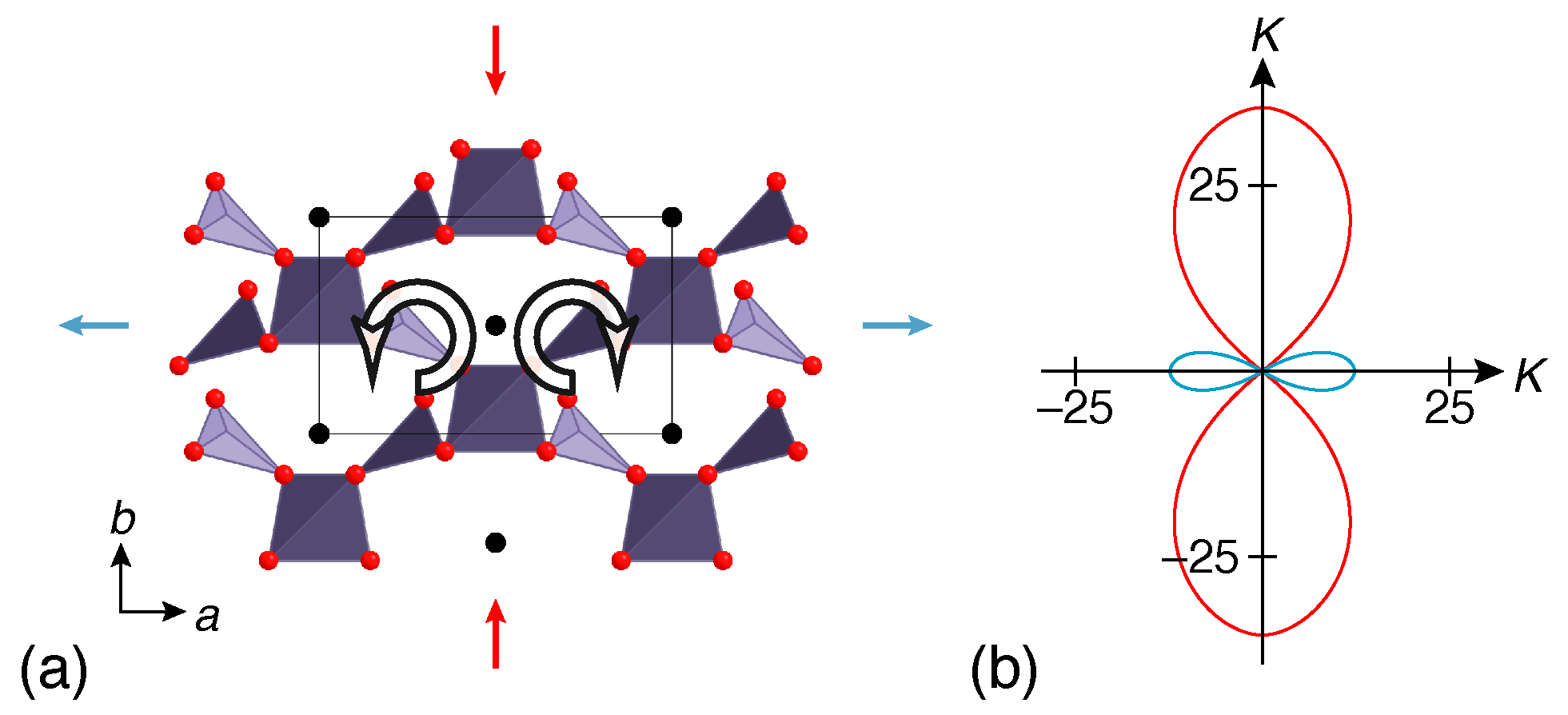}
\caption{(a) Correlated tilts of BO$_3$ units polyhedra lead to hinging of the borate framework in BiB$_3$O$_6$ and volume reduction; the network expands along $\mathbf a$ in the process.\cite{Dinnebier:2009} Bi atoms are shown as isolated black circles and the BO$_2$ network in polyhedral representation. (b) The compressibility indicatrix\cite{Cliffe:2012} determined from RUS measurements shown in the same orientation as (a): red and blue regions indicate, respectively, positive and negative values of the linear compressibility.\cite{Haussuhl:2006}}
\label{fig7}
\end{center}
\end{figure}

The concept that correlated polyhedral tilts might give rise to NLC even in the absence of a phase transition is by no means new---and is actually more likely to yield practically useful systems since NLC is then an intrinsic property of the ambient phase. In the nonlinear optic (NLO) material BiB$_3$O$_6$, for example, the dominant deformation mechanism under hydrostatic pressure involves correlated tilts of BO$_3$ units which act to hinge the connected borate framework.\cite{Dinnebier:2009,Haussuhl:2006,Stein:2007} This mechanism drives a strong contraction in one direction (parallel to the highly-compressible Bi$^{3+}$ lone pairs) that couples to a moderate expansion in a perpendicular direction [Fig.~\ref{fig7}(a)].\cite{Dinnebier:2009} There is no change in crystal symmetry throughout this process; in particular the mechanism cannot be thought of as arising in the vicinity of a ferroelastic phase transition.

A further point of interest regarding BiB$_3$O$_6$ is that it is one of the few NLC systems for which compressibilities have been determined using two complementary techniques. On the one hand, a variable-pressure crystallographic study reported $K_{\rm{NLC}}=-6.7(3)$\,TPa$^{-1}$ over the pressure range 0--5\,GPa (\emph{i.e.}, $\chi_K=3.35(15)$\%).\cite{Dinnebier:2009} On the other hand, experimental determination of the elastic tensor based on RUS measurements gave $K_{\rm{NLC}}=-12.5$\,TPa$^{-1}$, which is relevant in the limit $p\rightarrow0$ [Fig.~\ref{fig7}(b)].\cite{Haussuhl:2006} The factor-of-two difference between the two measurements of $K_{\rm{NLC}}$ does not imply an experimental inconsistency, but rather reflects the real variation in $K$ with pressure as illustrated in Fig.~\ref{fig2}. That NLC is relatively strong and persistent in BiB$_3$O$_6$ likely reflects the openness of its framework structure and the polarisability of the extra-framework Bi$^{3+}$ cation.

Equally strong NLC effects are expected to occur amongst other open framework structures based on connected polyhedra. In a recent computational study of 121 siliceous zeolites, a total of 16 were identified as NLC candidates on the basis of their calculated elastic compliance tensors.\cite{Coudert:2013} Amongst these frameworks, the strongest NLC behaviour is anticipated for the experimentally-realisable GIS zeolite topology,\cite{deBoer:1995} with $K_{\rm{NLC}}=-13.7$\,TPa$^{-1}$ in the $p\rightarrow0$ limit. What is remarkable here is the unexpected frequency of NLC: if the phenomenon occurs in 13\% of a broad family of materials then it is certainly less rare than originally envisaged.\cite{Coudert:2013} The pressure range over which these zeolites exhibit NLC remains to be determined; likewise the effects of Al substitution, extra-framework counterion inclusion, and solvation on NLC offer additional avenues of experimental investigation.

\begin{figure}[t]
\begin{center}
\includegraphics{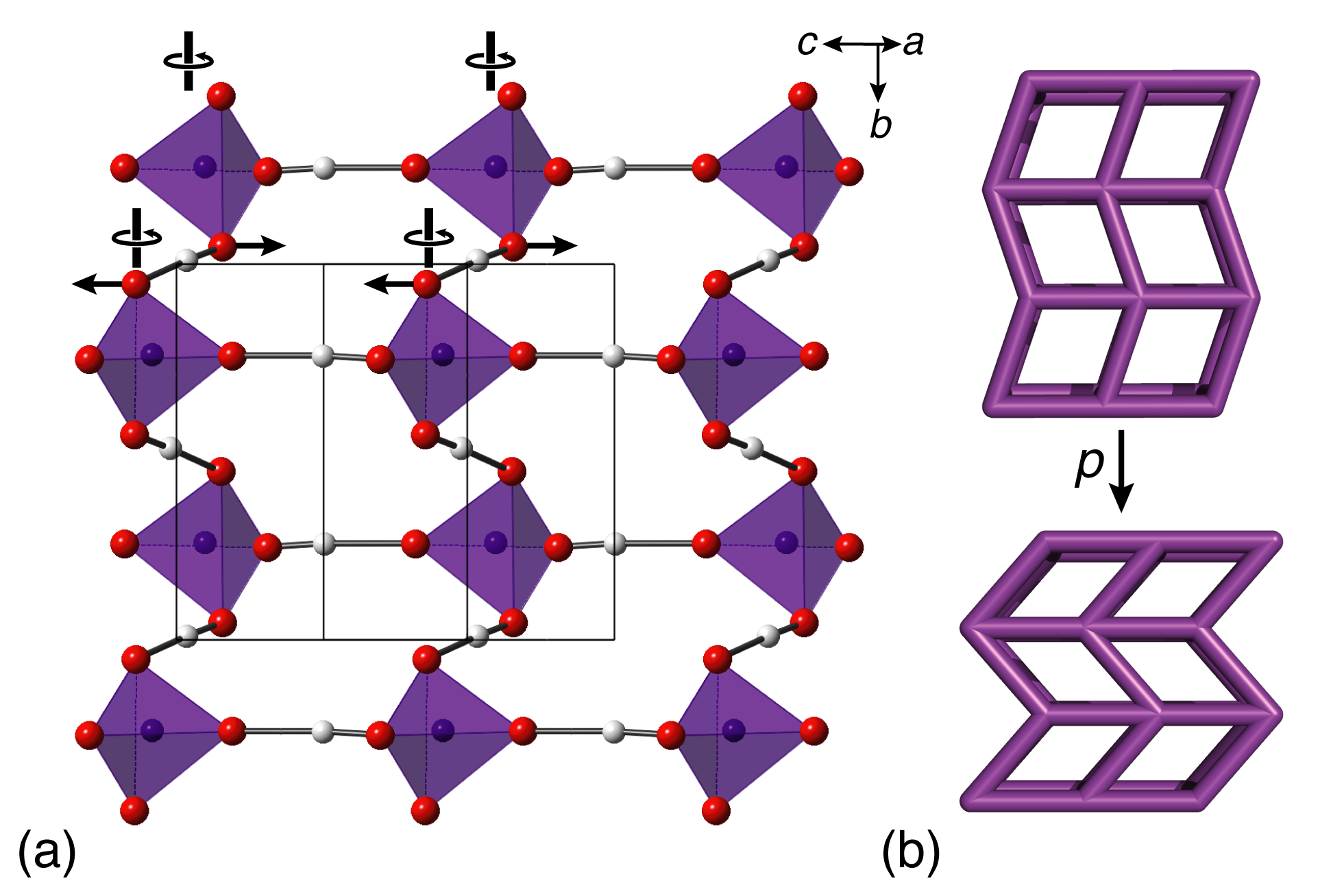}
\caption{(a) Correlated rotations of PO$_4$ tetrahedra about the $\mathbf b$ axis would cause the CsH$_2$PO$_4$ structure to densify while expanding along a direction parallel to $\mathbf c$.\cite{Prawer:1985} H atoms are shown as white spheres, PO$_4$ units in polyhedral representation, and Cs atoms omitted for clarity.\cite{Uesu:1976} (b) The densification mechanism of a generic herringbone lattice, which acts as a geometric model for NLC in CsH$_2$PO$_4$.}
\label{fig8}
\end{center}
\end{figure}

\begin{table*}
\small
  \caption{\ Compressibilities of materials for which NLC arises as a consequence of correlated polyhedral tilts.}
  \label{table2}
  \begin{tabular*}{\textwidth}{@{\extracolsep{\fill}}llllllll}
    \hline
  				& $K_{\textrm{NLC}}$ (TPa$^{-1}$)		& $K_{\textrm{PLC}}$ (TPa$^{-1}$)		& $\bar B$ (GPa) 	& Range (GPa) & $\chi_K$ (\%)	& Ref. \\ \hline
BiB$_3$O$_6$		& $-$6.7(3)						&--							&		  	& 0--5.0		& 3.35(15)		& \citenum{Haussuhl:2006} \\
CsH$_2$PO$_4$	& $-$260							& --							&			& 0			& --			& \citenum{Prawer:1985}  \\
BPO$_4$ 			& $-$0.92(10)						& 2.83(12)							&229(9)		&  0--56 		& 5.5(5)		& \citenum{Haines:2003} \\
BAsO$_4$ 		& $-$1.48(15)  						& 3.64(11)							&181(4)	  	&0--51 	& 6.2(7)  		& \citenum{Haines:2003} \\ \hline
  
   \end{tabular*}
\end{table*}
Correlated tilts of PO$_4$ tetrahedra have also been implicated in the strongly anisotropic elastic response of the widely-studied ferroelectric material CsH$_2$PO$_4$ [Fig.~\ref{fig8}(a)].\cite{Prawer:1985,Kobayashi:2003} To the best of our knowledge, the only variable-pressure crystallographic measurements of the ambient phase of this material are limited to (i) single-crystal neutron diffraction studies of the evolution of diffuse scattering,\cite{Youngblood:1980} and (ii) low resolution X-ray powder diffraction patterns at 0.29 and 2.89\,GPa reported without any subsequent structural analysis.\cite{Kobayashi:2003} To some extent this paucity of crystallographic data is surprising given the intense interest in the high-pressure ferroelectric behaviour of this material.\cite{Nelmes:1978} In the context of NLC it is perhaps even more remarkable because ultrasonic velocity measurements suggest a $p\rightarrow0$ compressibility of $-260$\,TPa$^{-1}$ along a direction approximately aligned with the $\mathbf c$ axis of the monoclinic cell.\cite{Prawer:1985} The powder diffraction patterns of Ref.~\citenum{Kobayashi:2003} do not show any obvious evidence of extreme NLC, although re-measurement across a larger number of more finely-spaced pressure intervals would help settle the issue definitively. The most likely mechanism responsible for NLC in the system (as suggested in Ref.~\citenum{Prawer:1985}) would involve PO$_4$ rotation-driven collapse of the herringbone hydrogen-bonding network as illustrated in Fig.~\ref{fig8}(b).

\begin{figure}[t]
\begin{center}
\includegraphics{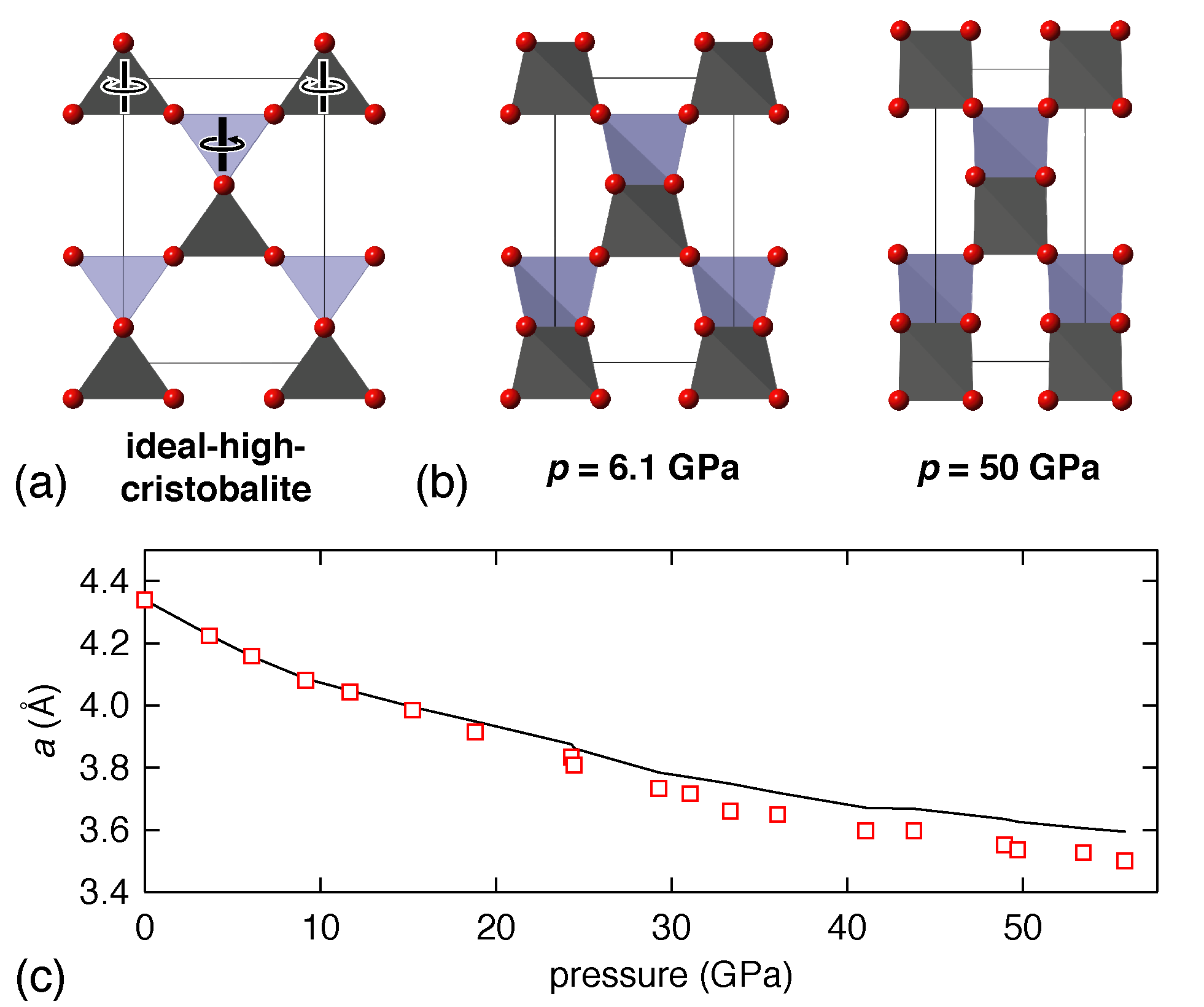}
\caption{(a) The structure of BEO$_4$ (E = P, As) compounds is related to that of cubic $\beta$-cristobalite \emph{via} decoration of the lattice with alternating BO$_4$ and EO$_4$ tetrahedra and subsequent rotation of all tetrahedral units about the tetrad axis by an arbitrary angle $0<\phi<\pi/4$.\cite{OKeeffe:1976,Leger:2001} (b) The structures of BPO$_4$ at high pressures are described by increasingly large values of $\phi$.\cite{Haines:2003} (c) That variations in $\phi$ dominate the compression mechanism is evident from a comparison of the actual variation in the lattice parameter $a$ (open squares) with that obtained from the variation in $\phi$ via Eq.~\eqref{phieq} (solid line).}
\label{fig9}
\end{center}
\end{figure}

Because extended structures tend to be less dense than their tilted counterparts, more often than not it will be the case that pressure-driven activation of tilt systems will favour PLC rather than NLC. Cristobalite-like BEO$_4$ (E = P, As) frameworks are an interesting case of \emph{indirect} NLC materials where NLC arises because a tilt-driven PLC mechanism results in a more rapid compression of the crystal lattice than can supported by the bulk material stiffness.\cite{Haines:2003} We proceed to explain this mechanism in more detail. Both BPO$_4$ and BAsO$_4$ adopt the same tetragonal $I\bar4$ variant of the cristobalite structure [Fig.~\ref{fig9}].\cite{OKeeffe:1976,Leger:2001} The dominant deformation mechanism involves correlated tilting of the BO$_4$/EO$_4$ tetrahedra around the tetragonal axis. The tilt angle $\phi$ is directly related to the $a$ lattice parameter via the projection of the mean B--O/E--O bond length onto the $(a,b)$ plane $r_\perp$:\cite{Haines:2003}
\begin{equation}\label{phieq}
a^2=\frac{16r_\perp^2}{\tan^2\phi+1}.
\end{equation}
Because bending of the B--O--E bonds carries a lower energy penalty than compression of the B--O or E--O bonds, the behaviour of the $a$ lattice parameter on compression is dominated almost exclusively by changes in $\phi$. What this means is that $K_a$ is effectively a measure only of the bending stiffness of the B--O--E linkages. In contrast, the bulk modulus $B$ measures the resistance to compression of the whole oxide lattice, which will depend largely on anion repulsion in three dimensions. For both BPO$_4$ and BAsO$_4$ the average value of this bulk modulus over the pressure range studied ($\bar B$) is sufficiently large that the inequality
\begin{equation}
K_c=\frac{1}{\bar B}-2K_a<0
\end{equation}
holds and NLC is observed along the tetragonal axis [Table~\ref{table2}].\cite{Haines:2003} More remarkable than the magnitude of NLC in these systems is the pressure range over which it is observed: Ref.~\citenum{Haines:2003} reports $K_{\rm NLC}=-0.92(10)$ and $-1.48(15)$\,TPa$^{-1}$ for E = P and As, respectively, over a pressure range of 0--52(5)\,GPa.

\subsection{Helices}

In many ways the mechanisms covered above---ferroelastic instabilities and polyhedral tilting---are similar to those often invoked in descriptions of other mechanical anomalies such as NTE.\cite{Barrera:2005} Over the next two sections our focus shifts away from these dynamical mechanisms to a consideration of \emph{topological} motifs that show a similar predisposition towards NLC.

Nature herself favours one particular geometric motif---namely, the helix---as a mechanism of generating and exploiting NLC in muscular response.\cite{Kier:1985} Just as NLC materials expand under hydrostatic pressure, so do they contract under negative (\emph{i.e.}, internal) pressure. When filled with fluid, helices exhibit precisely this response: their length decreases as the helix cross-section increases. Helical arrays of tendons enable muscle-like contraction that is driven by fluid injection rather than electrical impulse, and such a mechanism is implicated in the movement of certain types of worms, squid, and ancient limbless tetrapods.\cite{Kier:1985} The NLC community has long been aware of the related implication that non-biological NLC materials might be exploitable as artificial muscles and actuators if these motifs can be incorporated as part of materials design.\cite{Aliev:2009,Baughman:2003,Baughman:1998a,Foroughi:2011,Spinks:2002}

\begin{figure}[t]
\begin{center}
\includegraphics{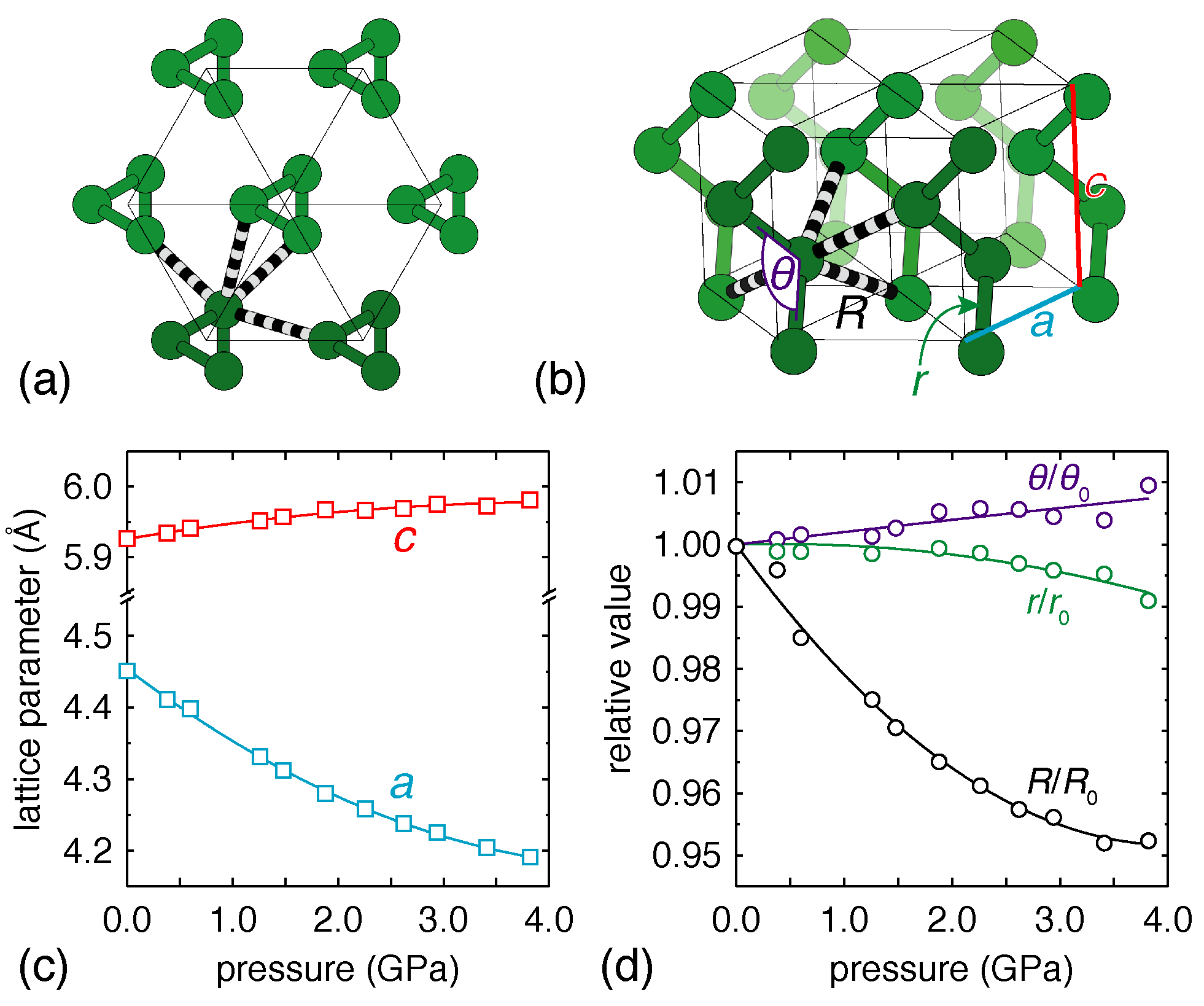}
\caption{(a) The crystal structure of selenium and tellurium consists of a triangular array of trigonal helices (shown in green). (b) The structure is completely described by the three parameters $r,R,\theta$ as described in the text. (c) Lattice parameter variation under hydrostatic pressure for Te, showing NLC parallel to the helix axis.\cite{Jamieson:1965, Bridgeman:1945} (d) The dominant compression mechanism involves reduction of inter-helix separation $R$ and unwinding of the helices (\emph{i.e.}, increasing $\theta$)---it is the latter that gives rise to NLC.}
\label{fig10}
\end{center}
\end{figure}

On the atomic scale, there are two remarkably simple chemical systems with helical structures that exhibit NLC under hydrostatic pressure. These are the trigonal polymorphs of elemental selenium and tellurium.\cite{Keller:1977} Both structures consist of a array of (enantiomorphic) trigonal helices packed on a triangular lattice [Fig.~\ref{fig10}]. Bonding interactions within any given helix are much stronger than those between helices so an interpretation of the bonding as molecular still has relevance despite the semiconducting properties of both systems. The densification required under increasing pressure can then be accommodated in two ways: either through compression of the weaker bonds between helices---which acts to decrease $a$ and leaves $c$ unchanged---or \emph{via} an increase in the `pitch' of each helix, compressing $a$ at the expense of some expansion along $c$. That both mechanisms operate for Se and Te is evident from variable-pressure crystallographic measurements [Fig.~\ref{fig10}].\cite{McCann:1972, McCann:1972a} It is conventional to reparamaterise the three measurable structural parameters---$a, c$, and $u$, the single free positional variable for Se/Te---in terms of the geometric parameters
\begin{eqnarray}
r&=&\left[3(ua)^2+\frac{1}{9}c^2\right]^{1/2},\\
R&=&\left[a^2(1-3u)+r^2\right]^{1/2},\\
\theta&=&\left[2\cos^{-1}(3ua/2r)\right].
\end{eqnarray}
Here $r$, $R$, and $\theta$ correspond, respectively, to the strongly-bonded E--E distance, the nearest weakly-bonded E--E distance, and the E--E--E angle within a strongly bonded helix.\cite{Keller:1977} Experimentally one finds that $R$ decreases most rapidly on increasing pressure: this reflects the rod packing compression mechanism described first above. In contrast, $r$ is essentially constant, meaning that the length of the helical path is essentially incompressible. By themselves, these two observations would not imply NLC. Instead the weak NLC behaviour that does occur [Table~\ref{table2}] can only be a consequence of the gentle increase in bond angle $\theta$ at higher pressures---\emph{i.e.}, the same pitch-variation mechanism Nature uses in muscle contraction.

\begin{table}[b]
\small
  \caption{\ Compressibilities of helical NLC materials.}
  \label{table4}
  \begin{tabular*}{\columnwidth}{@{\extracolsep{\fill}}lllllll}
    \hline
	&$K_{a}$		&$K_{c}$		 	&Range	&$\chi_{_{K}}$	& Refs. \\
	& (TPa$^{-1}$)		& (TPa$^{-1}$)		&(GPa)	& 	 (\%) 			 \\ \hline
Se 	&  12.0(6) 			& $-$2.5(4) 		 	& 0--5.2	&	1.3(2)		&	\citenum{McCann:1972, McCann:1972a}\\
Te 	& 13.6(9) 			&  $-$1.8(3) 			& 0--4	&	0.72(10)		&	\citenum{Jamieson:1965, Bridgeman:1945} \\ \hline
  \end{tabular*}
\end{table}

\begin{figure*}
\begin{center}
\includegraphics{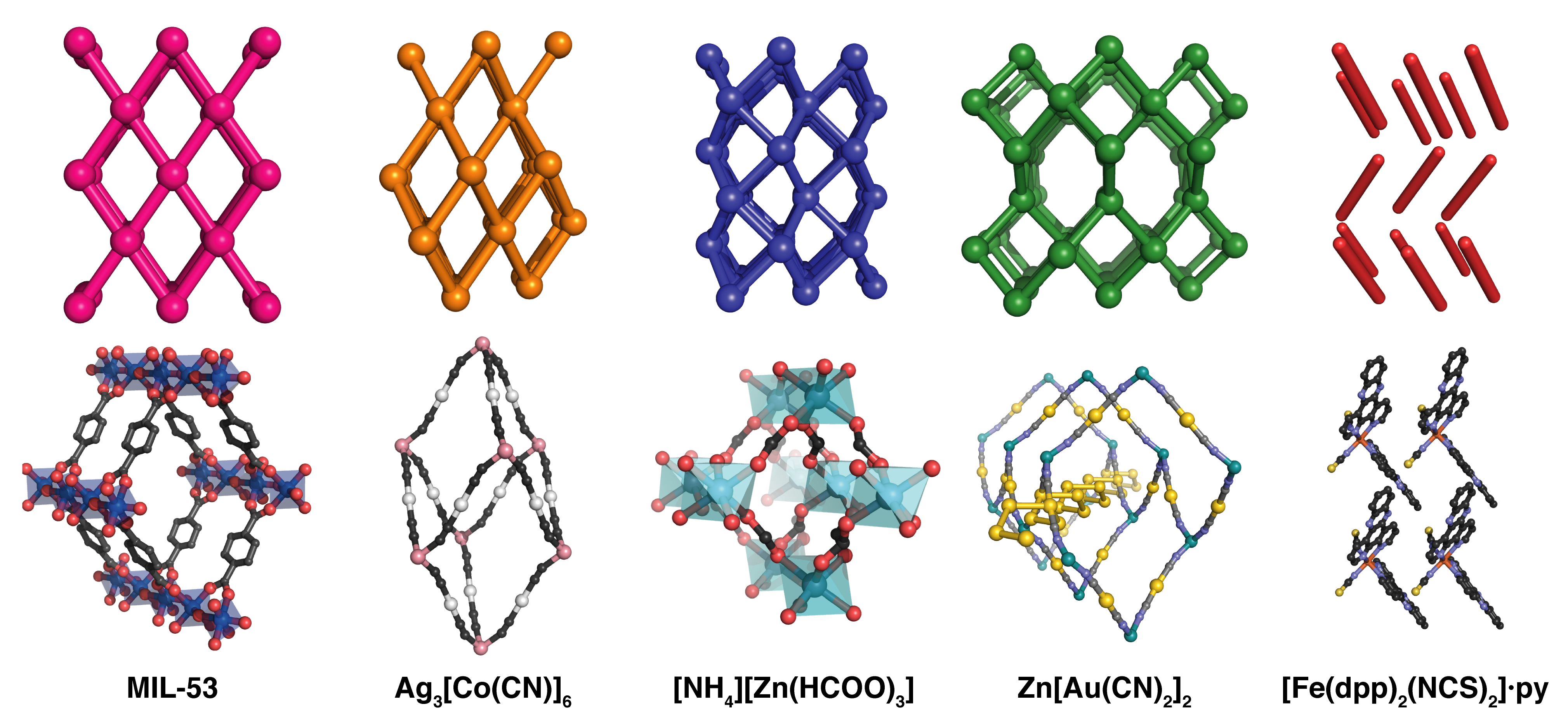}
\caption{Some wine-rack and honeycomb-like topologies known to favour NLC (top) and corresponding chemical systems (bottom). In all cases the mechanism responsible for NLC is analogous: densification involves extension of the lattice in the vertical direction.}
\label{fig11}
\end{center}
\end{figure*}

Remarkably, this NLC mechanism persists even when Se chains are incorporated within the cavities of zeolite AlPO$_4$-5 single crystals.\cite{Ren:2009} Interpretation of Raman spectroscopy and optical absorption measurements, together with {\it ab initio} calculations, converges on the same conclusion that the confined chains of Se do actually elongate under hydrostatic pressure.\cite{Ren:2009} Indeed, one anticipates that NLC may actually be quite a general phenomenon for ordered structures with helical motifs. It has been suggested elsewhere that the filamentous supramolecular polymer \{Au[(C$_2$H$_5$)$_2$NCS$_2$]\}$_n\cdot x$CH$_2$Cl$_2$ may exhibit negative compressibility \emph{via} such a mechanism;\cite{Paliwoda:2014} likewise helical metal--organic frameworks such as silver(I) dicyanamide (Refs.~\citenum{Britton:1990,Britton:1977}) may be interesting candidates for further investigation.


\subsection{Molecular Frameworks}

From an engineering perspective---rather than a biological one---the two geometric motifs most frequently associated with negative compressibility are the wine-rack and honeycomb networks (which are of course related to each other; see Fig.~\ref{fig11}).\cite{Grima:2012,Grima:2013} Both are characterised by extreme mechanical anisotropy and both have the property that their volumes are reduced under uniaxial expansion. Molecular framework chemistry offers an attractive means of designing materials that contain these same geometric features on the atomic scale.\cite{DelgadoFriedrichs:2007,OKeeffe:2009} Not only does the chemist have control over network topology, but the use of molecular linkers in the construction of framework materials leads to systems for which mechanical deformation mechanisms are dominated by framework flexing.\cite{Goodwin:2006} It is this feature that enables the mapping between chemical and engineering systems.

Within the field of metal--organic framework chemistry, the family of materials with a crystal structure most obviously resembling the wine-rack topology is probably the so-called MIL-53 system.\cite{Millange:2002} NLC behaviour was predicted for this family from both first principles\cite{Ortiz:2012,Ortiz:2013} and simple empirical\cite{Ogborn:2012} considerations; the first experimental verification of this behaviour has only recently appeared in print.\cite{SerraCrespo:2015} The extreme magnitude of NLC---Ref.~\citenum{SerraCrespo:2015} reports a compressibility value of $K_{\rm{NLC}}=-27$\,TPa$^{-1}$ over 0--2\,GPa---reflects the shallow energy potential that accompanies framework deformation for these systems and which is implicated in the well-known ``breathing'' effect on guest sorption.\cite{Serre:2002} Within the field there is probably an expectation that such large values are likely a general result for molecular frameworks, arising as a consequence of their low-density structures and the low energies of the supramolecular interactions affected by framework flexing.

Our own entry point into this field was \emph{via} the equally remarkable material silver(I) hexacyanocobaltate(III), Ag$_3$[Co(CN)$_6$].\cite{Goodwin:2008} Its trigonal structure can be considered a three-dimensional wine-rack and is topologically equivalent to three interpenetrating $\alpha$-Po (cubic) nets.\cite{Pauling:1968} Originally studied for its bizarre thermal expansion behaviour, the material was shown to admit very large strains ($\sim10$\%) even under mild conditions.\cite{Goodwin:2008,Goodwin:2008a} An extremely strong NLC effect ($K_{\rm{NLC}}=-76(9)$\,TPa$^{-1}$) was identified in a subsequent neutron scattering study.\cite{Goodwin:2008a} The mechanism responsible for NLC was straightforward enough: densification of the winerack-like framework proceeded via rapid compression of the $\mathbf a$ and $\mathbf b$ crystal axes and expansion along $\mathbf c$. Despite the magnitude of this NLC response, it is unlikely to find widespread application since in maximising $|K_{\rm{NLC}}|$ is framework becomes especially sensitive to shear instabilities. Indeed a shear-driven collapse occurs at $p=0.19$\,GPa, resulting in rapid densification and order-of-magnitude reduction in NLC behaviour.\cite{Goodwin:2008a} So while it seems that molecular frameworks can access much more extreme compressibilities than the ``conventional'' solid state materials described in the previous section, there is a clear challenge associated with balancing large flexibility with a propensity for mechanical instabiliity.

\begin{table*}
\small
  \caption{\ Compressibilities of NLC molecular frameworks and molecular solids.}
  \label{table3}
  \begin{tabular*}{\textwidth}{@{\extracolsep{\fill}}lllllllll}
    \hline
  												& $K_1$ (TPa$^{-1}$)	& $K_2$ (TPa$^{-1}$)& $K_3$ (TPa$^{-1}$)& $B_0$ (GPa) & Range (GPa) & $\chi_{_{K}}$ (\%)	& Ref(s). \\ \hline
Ag(mim) 											& $-$4.32(10) 			& 25.8(10) 	& 55(4)  		& 6.0(16) 			& 0--1.0		& 0.432(10) 				& \citenum{Ogborn:2012}\\
KMn[Ag(CN)$_2$]$_3$ 								& $-$12.0(8) 			& 33.2(13) 	&  33.2(13) 	& 12.7(11) 	 	& 0--2.2 		& 2.64(18)					& \citenum{Cairns:2012, Kamali:2013}\\
Ag$_3$[Co(CN)$_6$]-I								& $-$76(9) 			& 115(8) 		&  115(8) 		& 6.5(3) 		  	& 0--0.19 		& 1.44(17)					& \citenum{Goodwin:2008a}\\
Ag$_3$[Co(CN)$_6$]-II 								& $-$5.3(3) 			& 9.6(5) 		&  15.2(9)  	& 11.8(7) 		 	& 0.19--7.65 	&  4.0(2)					& \citenum{Goodwin:2008a}\\
Zn[Au(CN)$_2$]$_2$-I 								& $-$42(5) 			& 52(6) 		&  52(6) 		& 16.7(16) 		& 0--1.8 		& 7.6(9)					& \citenum{Cairns:2013a}\\
Zn[Au(CN)$_2$]$_2$-II 								& $-$6(3) 				& 16(5) 		& 16(5)   		& 27(3) 		  	& 1.8--14.2 	& 7(4)					& \citenum{Cairns:2013a}\\
\ce{[NH$_4$][Zn(HCOO)$_3$]} 						&  $-$1.8(8) 			& 15.8(9)		& 15.8(9) 		& 32.8(16) 	 	& 0--0.93 		& 0.17(7)					& \citenum{Li:2012}\\
ZAG-4 											&  $-$2.6(15)			& 7.9(5) 		&  29(3)  	& 11.66 		 	& 1.65--5.69 	& 1.1(6)   					& \citenum{Gagnon:2013}\\ \hline
[Fe(dpp)$_2$(NCS)$_2$]$\cdot$py 						&  $-$10(2) 			& 12(3) 		& 53(4) 		& 12.9(6) 		 	& 0--2.48		& 2.5(5)					& \citenum{Shepherd:2012} \\
CH$_3$OH$\cdot$H$_2$O 							&  $-$2.7(18) 			& 31.9(4)   	& 108.0(9) 	& 3.79(6) 		  	& 0--0.6 		& 0.16(11)					& \citenum{Fortes:2011} \\
{[(C$_6$F$_5$Au)$_2$($\mu$-1,4-diisocyanobenzene)]} 		&  $-$13(3) 			&  29(3)		& 31.3(4) 		& 7.5(7) 		 	& 0--2.42 		& 3.1(7) 					& \citenum{Woodall:2013} \\

    \hline
  \end{tabular*}
\end{table*}

One strategy for extending the range over which NLC is observed is to frustrate collapse by inclusion of counterions within the framework cavities.\cite{Cairns:2012} In the case of Ag$_3$[Co(CN)$_6$], for example, this can be achieved by substitution of Mn$^{2+}$ for Co$^{3+}$ while balancing charge with extra-framework K$^+$ ions.\cite{Geiser:2003} The resulting compound KMn[Ag(CN)$_2$]$_3$ remains mechanically stable up to at least 2.2\,GPa (the highest pressure for which diffraction data have been measured for this system), while retaining respectable NLC properties in the process: $K_{\rm{NLC}}=-12.0(8)$\,TPa$^{-1}$ over the range 0--2.2\,GPa.\cite{Cairns:2012} Once again the NLC mechanism involves hinging of the materials' wine-rack-like structure. The hypothesis that counterion inclusion resulted in ``soft-mode frustration'' has recently been verified directly using a combination of Raman spectroscopy and first principles calculations,\cite{Kamali:2013} and bears similarity to the sorption-induced stiffening of metal--organic frameworks documented elsewhere.\cite{Zhao:2015}

Already some design rules for maximising NLC begin to emerge:
\begin{enumerate}
\item{Mechanical anisotropy is clearly key: NLC requires PLC in an orthogonal direction since volume compressibility must remain positive.}
\item{Maximising this positive compressibility will likely maximise NLC.}
\item{Dynamic instabilities will reduce the pressure range over which NLC can be observed: using rigid molecular linkers and occupying void space with extra-framework cations or sorbate molecules may help extend this stability range}
\item{The network topology should likely be related to the wine-rack or honeycomb motifs.}
\end{enumerate}

These design principles were used to identify the first of the so-called ``giant'' NLC compounds: zinc dicyanoaurate(I), Zn[Au(CN)$_2$]$_2$.\cite{Cairns:2013a} The term ``giant'' demarcates exceptionally strong NLC ($K_{\rm{NLC}}<-30$\,TPa$^{-1}$) that persists over an industrially relevant pressure range (at least 1\,GPa). The quartzlike structure of this particular material is at once both anisotropic and related to the honeycomb net, satisfying respectively the first and fourth principles and so explaining the basic driving force for NLC.\cite{Hoskins:1995} The structure also has very little free volume because six of these nets interpenetrate one another;\cite{Hoskins:1995} this addresses the third principle and accounts for the extended pressure range over which NLC is observed (0--14.2\,GPa). But what sets Zn[Au(CN)$_2$]$_2$ apart is the strength of NLC is in its ambient phase ($K_{\rm{NLC}}=-42(5)$\,TPa$^{-1}$ over 0--1.8\,GPa), which arises from the extreme compressibility of supramolecular helical ``springs'' allowing especially strong PLC in one set of directions. Flexing of the honeycomb-like framework translates this exceptionally strong PLC into an equally remarkable NLC effect along the $\mathbf c$ crystal axis. It is this use of a supramolecular helix that addresses the second design principle: just as a spring is more compressible than the steel from which it is made, so too are the linear compressibilities of Zn[Au(CN)$_2$]$_2$ more extreme than would otherwise be expected.

\begin{figure}[t]
\begin{center}
\includegraphics[width=8.3cm]{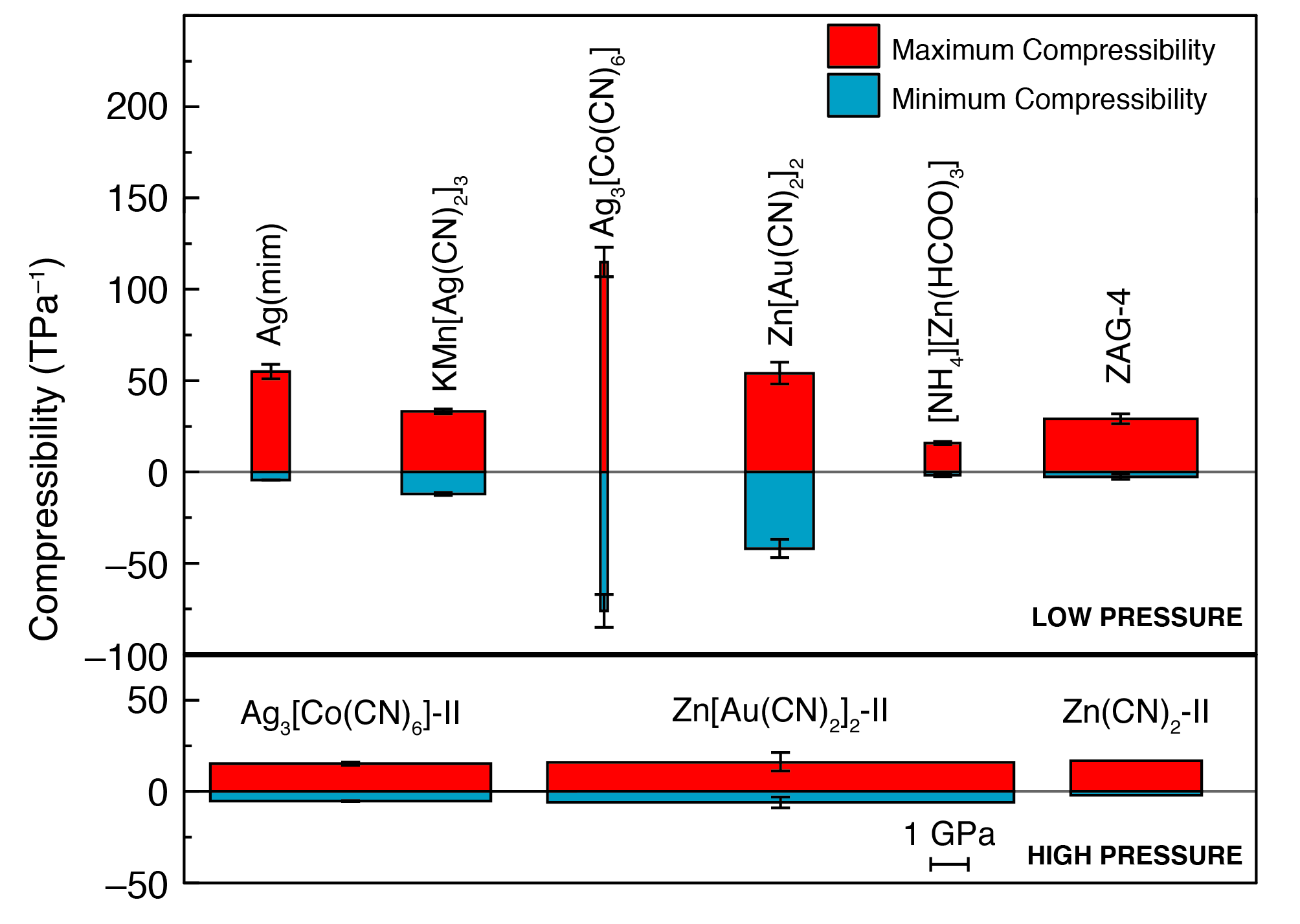}
\caption{Compressibility behaviour of a variety of molecular frameworks. The width of each bar is related to the pressure range over which compressibility is measured, and the height of each bar corresponds to the maximum (red) and minimum (blue) \emph{average} compressibility values over that range. Consequently the area of each bar is a measure of $\chi_K$: those materials with the greatest propensity to expand under pressure are those for which the area of the blue bar is largest.}
\label{fig12}
\end{center}
\end{figure}

NLC is increasingly frequently identified in a wide range of MOFs and molecular crystals beyond the various systems discussed above. Two relevant MOF examples are ammonium zinc(II) formate, [NH$_4$][Zn(HCOO)$_3$],\cite{Li:2012} and the zinc alkyl gate (ZAG) family.\cite{Gagnon:2013} The mechanism responsible for NLC in the former is essentially the same as that discussed for KMn[Ag(CN)$_2$]$_3$; all that differs between the two is their network topology (\emph{cag} \emph{vs} $\alpha$-Po).\cite{Li:2012} In contrast, the ``wine-rack'' NLC mechanism originally proposed for ZAGs (Ref.~\citenum{Gagnon:2013}) was later shown to be incorrect on the basis of quantum mechanical calculations.\cite{Coudert:2014} Instead it seems that NLC in this system is discontinuous and is driven instead by pressure-driven proton redistribution---a new mechanism altogether. The NLC behaviour of these MOFs is compared with that of other molecular frameworks in Table~\ref{table3} and Fig.~\ref{fig12}. In terms of molecular crystals, strong NLC effects can occur in situations where packing arrangements mimic the topological motifs known to favour NLC in framework structures.\cite{Goodwin:2010} So, for example, the molecular packing arrangements in systems as chemically diverse as methanol monohydrate, [Fe(dpp)$_2$(NCS)$_2$]$\cdot$py (dpp = dipyrido[3,2-$a$:2$^\prime$3$^\prime$-$c$]phenazine, py = pyridine) and [(C$_6$F$_5$Au)$_2$($\mu$-1,4-diisocyanobenzene)] are all related to the same wine-rack topology. Moderately strong NLC is observed in each case: $K_{\rm{NLC}}=-2.6(3), -10.3(20)$, and $-4.16$\,TPa$^{-1}$, for each of these three examples in turn.\cite{Fortes:2011,Shepherd:2012,Woodall:2013} 

\section{Discussion and Future Directions}

At the time of Baughman's original review (Ref.~\citenum{Baughman:1998}), NLC appeared to be a somewhat esoteric phenomenon that occurred in only a handful of peculiar systems. If the enumeration of various NLC materials and mechanisms given in our review leads to any one particular conclusion it is surely that NLC is rather more commonplace than might originally have been expected. From an experimental viewpoint, characterisation of NLC remains somewhat of a niche capability; however, the number of research groups with expertise in high-pressure crystallographic measurements is rapidly growing and we anticipate that NLC will be increasingly frequently observed as the scope of materials studied at high pressures diversifies. This is likely to be particularly true for molecular crystals, where the empirical predisposition towards herringbone-type packing arrangements will intrinsically favour NLC because of its relationship to the winerack topology.

But perhaps the most profound recent advance likely to influence our understanding of the true breadth of NLC behaviour is the relative ease and reliability with which first principles calculations can now determine elastic properties.\cite{Coudert:2013} This allows (relatively) rapid screening of entire classes of materials and pre-selection of the most interesting candidates for subsequent experimental investigation. We expect that the task of calculating elastic tensors for all known MOF structures, for example---once considered an inconceivably difficult process---will not remain computationally intractable for very much longer.\cite{Coudert:2014}

If NLC is likely to be discovered in a large number of new systems over the coming years, then what are the criteria for interesting and useful behaviour? Magnitude and range of NLC are two obvious metrics---hence the otherwise-questionable value of labels such as ``giant'' NLC.\cite{Cairns:2013a} But one other aspect of NLC behaviour deserves brief discussion: namely the relationship between NLC and crystal symmetry. Low symmetry intrinsically favours anisotropy---and hence NLC---since there are fewer constraints on the elastic stiffness tensor $\mathbf C$. While NLC is likely to be more prevalent in low-symmetry materials, the direction along which NLC occurs is less likely to have a fixed relationship to the crystal axes (hence bulk sample morphology) and moreover this direction itself will vary with pressure.\cite{Cliffe:2012} This matters because practical implementation of NLC requires careful material alignment. So, in general, we anticipate that NLC behaviour in uniaxial or orthorhombic systems will find greatest application because in these cases macroscopic alignment of the NLC axes can be assured.

\begin{figure}
\begin{center}
\includegraphics{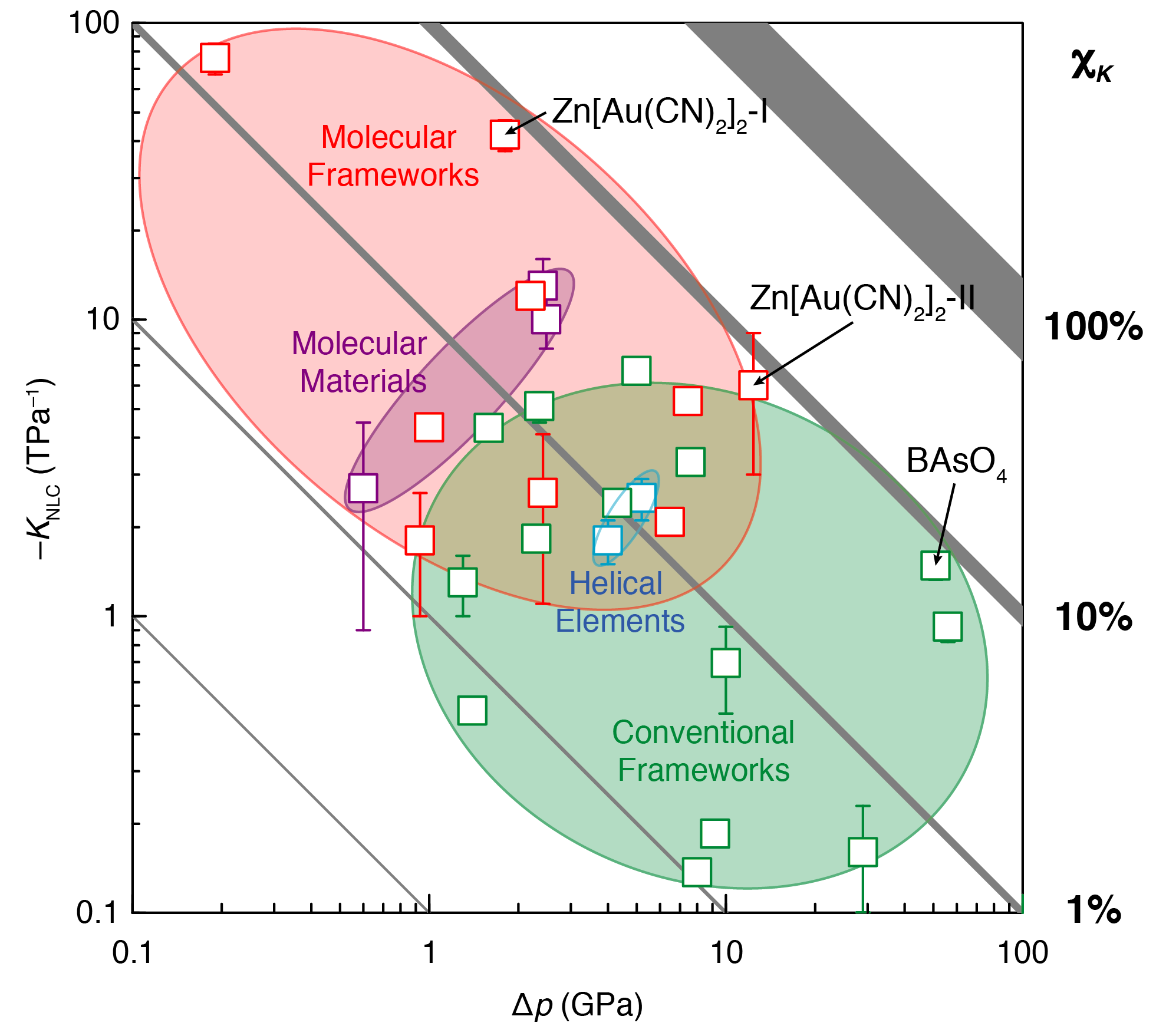}
\caption{Ashby-type diagram relating magnitude of NLC behaviour to pressure range over which it is observed for those materials included in this review. Diagonals correspond to points of constant compressibility capacity.}
\label{fig13}
\end{center}
\end{figure}

Having documented NLC in large variety of different chemical families, we sought to establish whether any universal trends in behaviour might emerge. Our approach is to compare for these different materials the relationship between magnitude of NLC effect and pressure range over which the phenomenon is observed. Our analysis is summarised in the Ashby-type diagram given in Fig.~\ref{fig13}. The diagonals in this representation correspond to points of constant $\chi_K$. What becomes obvious is that most NLC materials are distributed around the diagonal corresponding to $\chi_K=1\%$---this value is probably special only in the sense that it is large enough for NLC to have been noticed experimentally, but is not so large as to be extremely rare. Along this diagonal, those materials for which the mechanically-relevant bonding interactions are strong---the rutiles and framework silicates, for example---are clustered in a region where NLC pressure range is large but $K_{\rm{NLC}}$ itself is not so very extreme. In contrast, the flexible, open structures of MOFs and molecular frameworks cluster in a region where NLC is very strong but persists only over small pressure ranges.  In this way, eventual materials selection will be informed by whether range or magnitude is more critical for a given particular application of NLC. In terms of new materials discovery, the most attractive region of the Ashby plot is of course the (currently empty) top-right corner, where both metrics are maximised. Somehow the materials that exist in this region---if indeed they exist at all---must balance the weak interactions needed to produce strong NLC with the structural integrity required to avoid collapse at high pressures.

\begin{figure}[b]
\begin{center}
\includegraphics{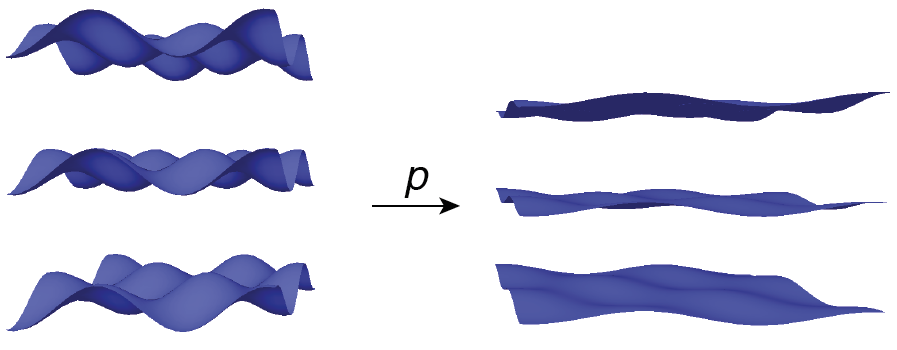}
\caption{Lifshitz mechanism for NAC.\cite{Lifshitz:1952} Densification of layered materials usually proceeds \emph{via} collapse in the stacking direction, which in turn results in an expansion in the two perpendicular directions (\emph{i.e.}, within the layer).}
\label{fig14}
\end{center}
\end{figure}

Our focus in this review has been almost entirely on materials which exhibit negative compressibility along just one principal axis. Negative area compressibility (NAC) is a natural extension and indeed modest NAC effects have been observed or predicted to occur in a handful of layered materials: silver(I) tricyanomethanide,\cite{Hodgson:2014} sodium vanadate,\cite{Loa:1999, Loa:1999a, Nakao:1998, Bernert:2001, Ohwada:1999} and TlGaSe$_2$ (Ref.~\citenum{Seyidov:2010}) (Table~\ref{table5}; we note that reports of NAC in PbTiO$_3$ remain contentious and so are not included\cite{Nelmes:1986,Huang:2012,Sani:2002}).  In each of these cases, the mechanism responsible for NAC is related to the so-called Lifshitz mode [Fig.~\ref{fig14}].\cite{Lifshitz:1952} An alternate mechanism---in which specific geometries of the various wine-rack and honeycomb-like topologies identified above couple densification to area expansion---has recently been proposed;\cite{Collings:2014} the extension to specific helical geometries also follows. Whatever the mechanism, it is reasonable to expect that NAC is unlikely ever to be as strong an effect as NLC. This is because the mechanical stability criterion of positive volume compressibility implies that the PLC effect along the axis perpendicular to the plane of negative compressibility must be at least twice as large as $K_{\rm{NAC}}$. In other words, NAC can only ever be half as strong as PLC.

\begin{table}
\small
  \caption{\ Compressibilities of NAC materials}
  \label{table5}
  \begin{tabular*}{\columnwidth}{@{\extracolsep{\fill}}lllll}
    \hline
&$K_{\rm{NAC}}$&$K_{\rm{PLC}}$&Range & Ref(s). \\
& (TPa$^{-1}$)& (TPa$^{-1}$)&(GPa)& \\\hline
Ag[C(CN)$_3$]&$-$7.5(8) & 66(20) & 0--0.615(6) & \citenum{Hodgson:2014}\\
NaV$_2$O$_5$&$-$3.128&1.67&2--10.0&\citenum{Loa:1999, Loa:1999a, Nakao:1998, Bernert:2001, Ohwada:1999}\\
TlGaSe$_2$ 		& $-$4.99 		& 33.28 		&--& \citenum{Seyidov:2010} \\
    \hline
  \end{tabular*}
\end{table}

While \emph{continuous} volume compressibility must be positive, there is no formal thermodynamic requirement forbidding volume increase across a pressure-induced phase transition. To the best of our knowledge, the only `realisation' of this bizarre phenomenon is a theoretical study of a fictitious multicomponent metamaterial assembled from cleverly-chosen pairwise potentials.\cite{Nicolaou:2012} Translating this study into real materials that exhibit a negative compressibility transition is difficult (perhaps impossible) for three main reasons. First, the expanded phase is only stable at high pressures in the 0\,K limit and is metastable at finite temperature. Second, the barrier to relaxation of this metastable state scales inversely with system size. For atomic-scale realisations--- \emph{i.e.} materials as discussed in this review---the barrier is unlikely to be sufficiently high to prevent rapid relaxation to the thermodynamic (compressed) state at high pressures.\cite{Nicolaou:2012} And, third, the pairwise potentials needed to produce the effect are as unusual as they are specific. Nevertheless, were it possible to realise even a \emph{discontinuous} negative compressibility transition in a chemical system then ``continuous'' negative volume compressibility behaviour might be feasible through inhomogeneous chemical doping to ``smear out'' the transition pressure---a trick used elsewhere to convert discontinuous thermal volume collapse to ``colossal'' NTE.\cite{Azuma:2011}

Whatever the particular manifestation of negative compressibility, what is clear is that its very existence is symptomatic of anomalies in the whole elastic tensor. One obvious area for development in the field involves the use of elastic property measurement techniques---historically applied in the fields of materials science, engineering, and metallurgy---as a means of exploring related elastic anomalies in the various chemical systems covered in this review. One expects that the general trends identified here will resurface irrespective of the particular measurement: strongly-bound materials will be more resistant to applied stress, and open frameworks will give rise to larger magnitude elastic responses. The development of new classes of negative Poisson ratio materials, for example, would be one such avenue of further research.\cite{Greaves:2011} 


From a materials chemistry perspective, however, perhaps the most exciting direction to be explored will be the coupling of anomalous elastic behaviour with other materials properties. The potential application of NLC in pressure sensing devices, for example, relies on the effect of simultaneous elongation and densification on optical behaviour.\cite{Baughman:1998} Orbital overlap in superconductors and magnetic materials---itself strongly sensitive to the distances between atoms---may also be tuneable in unexpected ways by exploiting the counterintuitive pressure dependencies attainable of NLC materials.\cite{Goh:2010} Thinking beyond NLC itself, the coupling of extreme elastic anisotropy (of which NLC is of course a simple consequence) with macroscopic dipole formation may lead to exciting new classes of piezoelectrics, ferroelectrics, and pyroelectrics. Whatever the future holds, the lessons learned from the study of NLC materials---namely the different ways in which chemical, geometric, and topological motifs influence elastic anomlies---are likely to play a crucial role in functional materials design.

\subsection*{Acknowledgments}

The authors gratefully acknowledge useful discussions with J.~A.~Hill and M.~J.~Cliffe (Oxford), I.~E.~Collings (Bayreuth), F.-X.~Coudert (Paris) and J.~Haines (Montpellier), and thank the EPSRC (grant EP/G004528/2) and ERC (project 279705) for financial support.

\balance 

\footnotesize{
\bibliography{pccp_2014_nlc} 
\bibliographystyle{rsc} 
}

\end{document}